\documentclass[apj,revtex4]{emulateapj}

\usepackage{amsmath}
\usepackage{lscape}
\usepackage{color}
\usepackage{graphicx}
\usepackage{hyperref}
\usepackage{multirow}

\newcommand{\Msolar}{M$_{\odot}\,$}

\begin{document}

\title{On the Origin of Sub-subgiant Stars. III. Formation Frequencies}
\shorttitle{Formation Frequencies of Sub-subgiant Stars}

\author{Aaron M.\ Geller$^{1,2,\dagger,}$\footnote{NSF Astronomy and Astrophysics Postdoctoral Fellow}, 
Emily M.\ Leiner$^3$,
Sourav Chatterjee$^1$,
Nathan W.\ C.\ Leigh$^{4}$,
Robert D.\ Mathieu$^3$,
Alison Sills$^5$}

\affil{$^1$Center for Interdisciplinary Exploration and Research in Astrophysics (CIERA) and Department of Physics \& Astronomy, Northwestern University, 2145 Sheridan Rd., Evanston, IL 60201, USA;\\
$^2$Adler Planetarium, Department of Astronomy, 1300 S. Lake Shore Drive, Chicago, IL 60605, USA;\\
$^3$Department of Astronomy, University of Wisconsin--Madison, 475 North Charter Street, Madison, WI 53706, USA;\\
$^4$Department of Astrophysics, American Museum of Natural History, Central Park West and 79th Street, New York, NY 10024, USA;\\
$^5$Department of Physics and Astronomy, McMaster University, Hamilton, ON L8S 4M1, Canada}

\email{$^\dagger$a-geller@northwestern.edu}

\shortauthors{Geller et al.}

\begin{abstract}

Sub-subgiants are a new class of stars that are optically redder than normal main-sequence stars and fainter than normal subgiant stars.
Sub-subgiants, and the possibly related red stragglers (which fall to the red of the giant branch), occupy a region of the 
color-magnitude diagram that is predicted to be devoid of stars by standard stellar evolution theory. 
In previous papers we presented the observed demographics of these sources and defined possible theoretical formation channels
through isolated binary evolution, the rapid stripping of a subgiant's envelope, and stellar collisions.
Sub-subgiants offer key tests for single- and binary-star evolution and stellar collision models.  
In this paper, we synthesize these findings to discuss the formation frequencies through each of these channels.
The empirical data, our analytic formation rate calculations, and analyses of sub-subgiants in a large grid of Monte Carlo globular cluster models
suggest that the binary evolution channels may be the most prevalent, though all channels appear to be viable 
routes to sub-subgiant creation (especially in higher-mass globular clusters).  Multiple 
formation channels may operate simultaneously to produce the observed sub-subgiant population.
Finally, many of these formation pathways can produce stars in both the sub-subgiant and red straggler (and blue straggler) regions of the 
color-magnitude diagram, in some cases as different stages along the same evolutionary sequence.  

\end{abstract}


\section{Introduction}
\label{s:intro}

\begin{figure}[!t]
\includegraphics[width=3.5in]{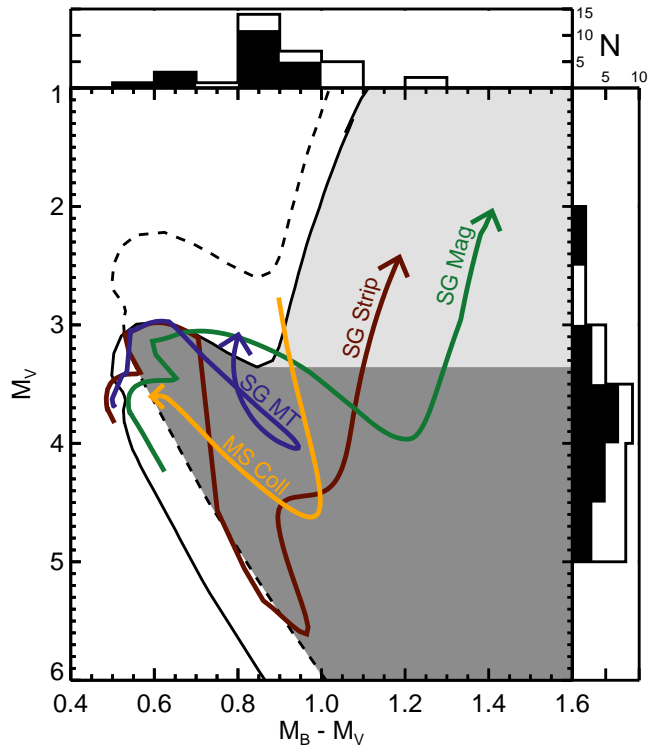}
\caption{
Color-magnitude diagram showing the theoretical SSG formation channels (colored lines) and our definition of the SSG region (dark gray shaded area) 
and RS region (light gray shaded area), with respect to a PARSEC isochrone \citep{bressan:12} for M67 (solid line).
In the main panel,
the dashed line shows the equal-mass binary sequence for M67.  
We show evolutionary tracks from MESA \citep{paxton:15} for a subgiant undergoing stable mass transfer (``SG MT'', purple line), a subgiant that has been stripped of much of its envelope (i.e., after ejecting
a common-envelope or after a grazing collision ``SG Strip'', dark-red line), and a star that has a reduced convective mixing length coefficient (and increased 
magnetic activity, ``SG Mag'', green line).  We also show the result of a collision between two 0.7~\Msolar\ MS stars (from \citealt{sills:02}
``MS Coll'', yellow line).  For all colored lines, the arrows indicate the direction of time along the evolutionary sequence.
In the two sub-panels, we show histograms of the observed distribution of SSG and RS stars from the open and globular clusters studied in Paper I, in ${\rm M_V}$ (right)
and ${\rm M_B - M_V}$ (top).  The black-filled histograms show the contribution from the globular cluster sources alone, with the additional white-filled region (up to the solid lines)
coming from the open cluster sources.  We refer the reader to Figure 1 in Paper 1 for CMDs of these stars in the individual clusters.
\label{f:CMDreg}
}
\end{figure}

This is the third paper in a series investigating the origins of a relatively new class of stars known as sub-subgiants (SSGs).  In \citet[][hereafter Paper I]{geller:17}, we gather the 
available observations for these sources and find that SSGs share the following important empirical characteristics:
\begin{enumerate}
\item SSGs occupy a unique location on a color-magnitude diagram (CMD), redward of the normal MS stars but fainter than the subgiant branch, where normal single-star evolution does not predict 
stars (see Figure~\ref{f:CMDreg}).
\item More than half of the SSGs are observed to be X-ray sources, with typical luminosities of order 10$^{30-31}$ erg s$^{-1}$, consistent with active binaries.
\item At least one third of the SSGs exhibit H$\alpha$ emission (an indicator of chromospheric activity).
\item At least two thirds of the SSGs are photometric and/or radial-velocity variables, with typical periods of $\lesssim$15 days.
\item At least three quarters of the variable SSGs are radial-velocity binaries.
\item The specific frequency of SSGs increases toward lower-mass star clusters.
\end{enumerate}
The fractions of sources given in items 2-5 above are all lower limits because not all sources have the necessary observations to investigate each characteristic.

In \citet[][hereafter Paper II]{leiner:17}, we study three specific formation channels in detail that can produce stars in the SSG region of the color-magnitude diagram (CMD),
namely 
(i) ongoing mass-transfer from a subgiant donor, ``SG MT'',
(ii) a reduced convective efficiency, likely related to increased magnetic activity, ``SG Mag'', and
(iii) rapid and partial stripping of a subgiants' envelope, ``SG Strip''.
Paper II also briefly considers a fourth channel, 
(iv) a main-sequence -- main-sequence stellar collision, where the collision product is observed while settling back down onto the normal main-sequence, ``MS Coll''.
We provide details and models for these mechanisms in Paper II (and references therein), and describe them qualitatively in Section~\ref{s:fchan} below.
For reference, in Figure~\ref{f:CMDreg} we show example evolutionary tracks for each of these mechanism, plotted over an isochrone at the age of M67.
Within this same figure we also define the SSG region
on a CMD, with the dark-gray shading (see also Figure 1 in Paper I).

The ``red straggler'' (RS) stars, which are possibly related to the SSGs, are found in the region with the light-gray shading in Figure~\ref{f:CMDreg}.  
We note here (and in Papers I and II) that there has been some confusion in the literature with the naming convention of these two types of stars.  We urge readers to adopt the convention that we 
set forth in this series of papers to identify SSG and RS stars.
There are far fewer RS stars than SSGs, but, despite their different location on the CMD,
their empirical characteristics appear to be very similar to the SSG stars.  
Some of the SSG formation channels discussed in Paper II predict an evolutionary relationship between stars in the SSG and RS regions (where one is the precursor to the other), and 
furthermore at least two of the formation channels (``SG MT'' and ``MS Coll'') can lead to the formation of a blue straggler star (BSS).

In this paper, we investigate the formation rates of SSGs through these four mechanisms through analytic calculations.  Our goal is to identify if indeed all of these mechanisms are viable,
or if one or more clearly dominates the production of SSGs.  After providing a qualitative description of the four formation 
mechanisms in Section~\ref{s:fchan},  we then discuss the probabilities of observing SSGs from each of these theoretical formation channels in Sections~\ref{s:pchan} and~\ref{s:prob}.  We 
investigate SSGs created in $N$-body and Monte Carlo star cluster models in Section~\ref{s:nbody}.  Finally, in Section~\ref{s:discuss} we provide a brief discussion and conclusions.

\section{Summary of Theoretical Formation Channels}
\label{s:fchan}

In Paper II we study SSG formation channels in detail, primarily through in-depth analyses of MESA models \citep{paxton:15}.
In this section, we provide a brief summary of these formation channels, specifically for SSGs formed through
ongoing binary mass transfer (Section~\ref{s:SGMT}), 
increased magnetic activity leading to inhibited convection (Section~\ref{s:SGMag}), 
rapid loss of an envelope (Section~\ref{s:SGStrip}), 
and MS -- MS collisions (Section~\ref{s:MSColl}).
We refer the reader to Paper II for further details about the physics behind the first three mechanism 
(and a wider exploration of parameter space).  MS stellar collisions have been modeled previously in detail 
 \citep[e.g.][]{sills:97,sills:01,sills:02,sills:05}, in the context of BSS formation.

\subsection{Ongoing Binary Mass Transfer Involving a Subgiant Star (``SG MT'')}
\label{s:SGMT}

If a binary containing a MS star has a short enough orbital period ($\sim$1.5 days for a 
circular binary with a primary star at the turnoff in M67), this MS star can overfill its Roche lobe shortly after evolving off of the MS.  
As mass transfer begins, the now subgiant star loses mass, becomes fainter and moves into the SSG region.  

The MESA model in Figure~\ref{f:CMDreg} shows a 1.0 day binary with a 1.3 \Msolar\ primary and a 0.7 \Msolar\ secondary. 
Mass transfer begins when the primary overflows its Roche lobe on the subgiant branch. Stable mass transfer proceeds with an efficiency of 50\%. 
In this model, the binary remains in the SSG region for $\sim$400 Myr, 
which is comparable to the duration of the subgiant phase of a normal $\sim$1.3 \Msolar\ star (of 600 Myr in MESA).

In this scenario, the subgiant must be the brighter star in the binary in the optical for it to appear in the SSG region.  
The accretor could be a MS star or a compact object.
However, if a MS accretor is massive enough initially,
it may gain enough mass to become a BSS,
and dominate the combined light,
before the subgiant becomes sub-luminous enough to enter the SSG region (unless the mass transfer is extremely nonconservative). 

Here, one would expect to observe a short-period binary, likely with a rapidly rotating primary (subgiant) star.  
Photometric variability could potentially arise from ellipsoidal variations on the subgiant, spot activity, or eclipses. 
X-rays (and H$\alpha$ emission) could be produced by chromospheric activity on the rapidly rotating subgiant and/or hot spots in the accretion stream onto a 
compact object (likely requiring a neutron star or black hole accretor to reach X-ray temperatures).

In many cases where the accretor is a MS star, the evolution will lead to the coalescence of the two stars, 
and interestingly, in this scenario, the merger product may be observed eventually as a BSS.
The mass-transfer model shown in Figure~\ref{f:CMDreg} eventually creates a BSS with a WD companion.
Indeed binary mass transfer is one of the primary BSS formation mechanisms \citep{mccrea:64,tian:06,chen:08,geller:11},
especially in low-density ($\rho_c\lesssim10^3 Msun/pc^3$) environments \citep{chatterjee:13a}.
We will come back to this potential connection between SSGs and BSS in Section~\ref{s:discuss}.

\subsection{Increased Magnetic Activity in a Subgiant Star (``SG Mag'')}
\label{s:SGMag}

While the effects of magnetic fields on stellar evolution are in general not well known, there is evidence that magnetic fields 
may alter the temperature and radii of stars by lowering the efficiency of convection. For example, low-mass eclipsing binaries are 
found to be larger and cooler than model predictions, which has been attributed to magnetic activity \citep[e.g.,][]{chabrier:07,clausen:09}. 
A similar mechanism may be at work in SSGs (Paper II).
\citet{chabrier:07} lower the mixing length coefficient in their models to mimic lowered convective efficiency.
Our preliminary MESA models, where we reduce the convective mixing length coefficient to $\alpha=1.2$ (see green line in Figure~\ref{f:CMDreg}, 
and note that MESA's standard mixing length coefficient is $\alpha=2.0$), 
suggest that this mechanism may produce SSGs primarily after the downturn on the subgiant branch and on the lower red-giant branch (RGB).

\subsection{Rapid Loss of a Subgiant Star's Envelope (``SG Strip'')}
\label{s:SGStrip}

If the envelope of a subgiant star is rapidly stripped away it will become fainter, while losing mass\footnote{\footnotesize Stripping the envelope of a
red-giant star has only a very minor affect on its luminosity \citep{leigh:16b}, because the luminosity of a red giant is controlled almost entirely by the 
He core.  Thus, despite the larger physical size of a red giant, and therefore the larger collision rate, this stripping mechanism may be most easily 
observed for subgiants.}.  
Once mass loss stops, the star will begin 
to evolve toward the subgiant and giant branches as before, but now along a path appropriate for its new lower mass.  If enough mass is lost, the star will be 
fainter than the cluster's subgiant branch and eventually redder than the giant branch, moving through the observed SSG region.  

For the ``SG Strip'' track shown in Figure~\ref{f:CMDreg}, we use MESA to evolve a 1.3 \Msolar\ star,
and we remove 0.45 \Msolar\ from the envelope soon after the star begins to evolve off of the MS and 
at a rate of 10$^{-5}$ \Msolar\ yr$^{-1}$. At this rate the star is driven out of thermal 
equilibrium initially, but quickly returns to an equilibrium position once the removal of mass is complete. (This is the largest mass transfer rate we are able 
to reliably model in MESA for this stellar mass and evolutionary state; at larger rates hydrodynamical effects become important.)

One possible method to induce this rapid mass loss is a grazing collision between a subgiant and some more compact star (perhaps a compact object or MS star), 
with an impact parameter small enough to strip the 
subgiant's envelope, but large enough that the two stars don't directly merge.  We will refer to this pathway as ``SG Coll''.
A second potential method is through the ejection of a common-envelope; we will refer to this pathway as ``SG CE''.
For our purposes here, we do not consider what happens to the mass lost from the subgiant (whether it can be accreted by the other star or lost from the system entirely).
Both of these mechanisms require further detailed modeling; here we will simply assume that they are both possible, and focus on the possibility of 
observing the product of such an event.

After both
processes,
a tight binary companion could remain (for the ``SG Coll'' scenario, this could be akin to a tidal capture; see e.g., \citealt{fabian:75} and \citealt{press:77}). 
The subgiant may be spun up in this process.  If the stripped subgiant is rotating rapidly, then one may expect to observe 
photometric variability and X-ray emission due to chromospheric activity and spots.  

\subsection{Collision of Two MS Stars (``MS Coll'')}
\label{s:MSColl}

In Figure~\ref{f:CMDreg} we show a collision product from \citet{sills:97} resulting from two 0.7 \Msolar\ stars.
Immediately after a collision between two MS stars, the collision product will become brighter (due primarily to the kinetic energy input from the 
motion of the stars leading up to collision) by a factor of about 10 to 50 (in luminosity) for the mass range of interest here.  Afterwards the star will settle back into 
thermal equilibrium by contracting and releasing gravitational potential energy, along analogous tracks to pre-MS stars.  Through this contraction phase, the star
becomes fainter and eventually settles back near the normal MS stars, but before reaching the MS the collision product may reside in the 
SSG region.  The contraction phase occurs over roughly a thermal timescale, which is between about 1-15 Myr for the masses of interest here.

If the collision is off axis, the product will likely be very rapidly rotating \citep{sills:05}, which could lead to similar photometric variability and X-ray emission 
as (particularly if a magnetic field can be maintained) as observed for some SSGs.  Scattering experiments and $N$-body star cluster simulations suggest that 
it would be difficult for the collision product to retain a binary companion at the short periods that are observed for many SSGs (i.e., of order 10 days) directly 
after a collision \citep[e.g.][]{fregeau:04,leigh:11,geller:13}. Subsequent exchanges or tidal capture encounters could become more likely with the increased mass 
(and temporary increase in radius) of the collision product. Further scattering experiments and $N$-body models are necessary to better understand
the likelihood for creating a short-period binary containing a collision product within such a short timescale after the collision (as would be required to produce 
SSGs in binaries with periods of order 10 days).

Though we show one specific collision model, a wide range of component masses can produce collision products in the SSG region.
Furthermore, for certain combinations of MS stars, the collision product may be ``born'' in the RS region and contract through the SSG region as it settles back into 
thermal equilibrium.  
This mechanism has also been invoked to explain BSS \citep[e.g.][]{hills:76, leonard:89, sills:09}.  
A collision product that could be observed as a SSG may later be observed as a BSS, after the normal stars of similar mass evolve toward the subgiant and giant 
branches.

\pagebreak
\section{Probabilities of Observing The Products of Each Formation Channel}
\label{s:pchan}

Each of these theoretical formation channels can produce products that have characteristics consistent with at least
a subset of the observed SSGs.  Many of these products are predicted to be relatively short-lived in relation to the age of the clusters that have 
SSGs.  We investigate here the probability of observing at least one SSG from each mechanism, respectively, in different star clusters, 
both over a range in cluster masses (e.g., Figure~\ref{f:poisson}) and for the observed parameters of the specific clusters that have SSGs (e.g., Table~\ref{probtab} and Figure~\ref{f:bar}). 

We follow the same framework in our calculations for each mechanism, based on the cumulative Poisson probability:
\begin{equation} \label{e:poisson}
\Psi(t,\tau) = 1 - e^{-\left(t/\tau\right)}\sum_{x=0}^{n-1} \frac{\left(t/\tau\right)^x}{x!} ,
\end{equation}
where $t$ is the time interval of interest (here, the duration that the star remains in the SSG region), 
$\tau$ is the mean time in between events, and $n$ is the number of events. 
Equation~\ref{e:poisson} gives the probability of observing $n$ or more events over the time interval $t$, when the 
mean number of events is expected to be $t/\tau$.  We discuss our estimates for $t$ and $\tau$ for each respective formation channel below, 
and in all cases we attempt to take the most optimistic assumptions.

First, our timescale calculations depend on the cluster age, mass ($M_\text{cl}$), 
metallicity ([Fe/H]), binary fraction ($f_\text{b}$), central velocity dispersion ($\sigma_0$), central density ($\rho_0$), core radius ($r_\text{c}$) and/or 
half-mass radius ($r_\text{hm}$).  We describe how we obtain these values in Section~\ref{s:prob}.
In general, for our study of the specific clusters (Section~\ref{s:pclu}), we obtain values from the literature (Table~\ref{probtab}). 
For our general calculations (Section~\ref{s:pgen}, and also as estimates for cluster specific values that are unavailable in the literature), 
we assume a \citet{plummer:11} model and also use the semi-analytic cluster evolution code EMACSS \citep{alexander:12,gieles:14,alexander:14}.

To start, we use the rapid Single Star Evolution code SSE \citep{hurley:00} to determine the mass of a star that would reside at the 
base of the giant branch for a given cluster age and metallicity.  
We take the evolutionary states for stars in these calculations directly from SSE.
We will refer to this star as $S_1$ below.  
We then use SSE to determined the mass, radius and luminosity of this star
when it was on the zero-age main sequence (ZAMS), the terminal-age MS (TAMS), and at the base of the RGB, for a given metallicity.  

For many of the scenarios, we also require the number of subgiants (or the fraction of stars that are subgiants, $f_\text{SG}$) expected to be in a given cluster.  To estimate this value,
we first determine an appropriate mass function of a cluster of a given age and mass using the method of \citet{webb:15}, 
which accounts for the change to a \citet{kroupa:01} IMF due to dynamical evolution and mass loss from the
cluster\footnote{The true cluster mass function depends on many uncertain factors (e.g., the IMF, initial Jacobi filling factor, remnant retention fractions, etc.) which are neglected in the simplified \citet{webb:15} relation. However, this simplified relation is sufficient for the approximate calculations performed here.}
This method requires an estimate of the initial cluster mass, which we derive by iteratively modeling clusters of different 
initial masses using EMACSS until reproducing the observed present day cluster mass 
(at either the solar Galactocentric distance, for Section~\ref{s:pgen}, or the true Galactocentric distance of the given cluster, for Section~\ref{s:pclu}).
We then use SSE to estimate the masses of stars that would evolve off the MS at +/- 1 Gyr from the cluster age.  These masses, combined with the 
mass function, provides a rate at which stars evolve off the MS at the given cluster age and metallicity, $\Gamma_\text{ev}$.  
This rate multiplied by the lifetime of $S_1$ on the subgiant branch yields an estimate of the number of subgiant stars in a given cluster 
(and a similar method can provide the number of MS stars in the cluster).

This theoretical estimate for the number of subgiants is consistent with observed values.  For instance, in the open clusters studied in Paper II, 
we count 20-30 subgiant stars in M67, and about 100 subgiant stars in NGC 6791.  (These numbers, of course, depend on where one defines the end of the MS and 
the base of the RGB, which can be somewhat subjective on a CMD.)
Following the theoretical procedure above, we predict 32 subgiants in M67 and 120 in NGC 6791, both consistent with the observed values.

Given the mass function, we can also estimate the mean single-star mass in the cluster, $\left<m_\text{s}\right>$.  For some calculations, we also desire the mean 
mass of an object (single or binary).  We estimate this value as 
$\left<m\right> = \left(1 - f_\text{b}\right) \left<m_\text{s}\right> + f_\text{b} \left<m_\text{b}\right>$, where 
$f_\text{b}$ is the cluster binary fraction, $\left<m_\text{b}\right>$ is the mean binary mass, and we assume a mean binary mass ratio of 0.5 (a reasonable guess for an 
approximately uniform mass ratio distribution, as is observed for solar-type binaries in the Galactic field
and globular clusters; see e.g., \citealt{raghavan:10} and \citealt{milone:12})
such that $ \left<m_\text{b}\right> =  1.5 \left<m_\text{s}\right>$.

For our general calculations, discussed in Section~\ref{s:pgen}, we obtain the binary frequency, $f_\text{b}$, for globular clusters from the empirical study of \citet{leigh:13}.  
For open clusters, we estimate $f_\text{b}$ by first assuming that prior to dynamical disruptions the binaries would follow the field solar-type stars, with 
a 50\% binary frequency and a log-normal binary period distribution 
\citep[][with a mean of $\log (P \text{[days]})~=~5.03$ and $\sigma = 2.28$]{raghavan:10}.  Then we truncated the period distribution
at the hard-soft boundary,
\begin{equation} \label{e:phs}
P_\text{hs} = \frac{\pi G}{\sqrt{2}} \left(\frac{m_1 \left<m_\text{s}\right>}{\left<m\right>}\right)^{3/2} \left(m_1+\left<m_\text{s}\right>\right)^{-1/2}\sigma_0^{-3} ,
\end{equation}
derived using the virial theorem to relate the mean binary binding energy to the local mean kinetic energy of a colliding star,
where $m_1$ is the initial mass of $S_1$, and $\sigma_0$ is the three-dimensional velocity dispersion in the core
(and we assume a \citet{plummer:11} model and that $\sigma_0=\sqrt{3}\sigma_\text{0,1D}$).
We calculate the cluster binary frequency as the ratio of the area under the truncated period distribution to that of the full distribution 
times the 50\% solar-type field binary frequency.
This assumes that the cluster has lived through sufficient relaxation times that all binaries have cycled through the core, which is 
reasonable for the open clusters known to contain SSGs.
(A more detailed calculation might account for the time and radial dependence of the hard-soft boundary, but that is beyond the scope of this paper.)
This produces binary fractions consistent with open cluster observations \citep[e.g.,][]{geller:12,geller:15}. 
In practice, this method for open clusters requires an iterative derivation of $f_\text{b}$, $\left<m\right>$ and $P_\text{hs}$.
For our cluster-specific calculations, discussed in Section~\ref{s:pclu}, we take the observed binary fractions (where available).

In the following, we describe our derivation of the timescale $\tau$ from Equation~\ref{e:poisson} for each specific formation mechanism. 
For the MS--MS collision channel, we also derive $t$, while for all others we simply take $t$ equal to the lifetime of $S_1$ on the subgiant branch.
Again, our assumption for $t$ represents the most optimistic scenario for the duration of each mechanism.

\subsection{Ongoing Binary Mass Transfer Involving a Subgiant Star}
\label{s:pSGMT}

We calculate $\tau$ here as the mean time between stars in appropriate binaries evolving off of the MS.  
Only binaries with orbital periods large enough to avoid
Roche lobe overflow (RLOF)
on the MS and small enough to undergo RLOF on the subgiant 
branch are of interest, which defines a fraction of the binary population by period $f_\text{P}$.  
Here, we use the Roche radius equation from \citet{eggleton:83}:
\begin{equation} \label{e:roche}
\frac{r_L}{a} = \frac{0.49 q^{-2/3}}{0.6 q^{-2/3} + \ln\left(1 + q^{-1/3}\right)},
\end{equation}
where we set $q = \left<m_\text{s}\right>/m_1$,
$a$ is the binary's semi-major axis, and we assume circular orbits (a standard assumption, given the expectation of tidal circularization,
and sufficient for these approximate calculations).
Likewise only
binaries expected to undergo stable mass transfer are of interest.  We impose a critical mass ratio of 
$q_\text{crit} = m_\text{accretor}/m_\text{donor} = 1/3$, below which we assume that the system undergoes a common 
envelope and is not included in this particular mechanism.
The value of 1/3 is similar to values used in binary population synthesis codes for such stars 
(e.g., \citealt{hurley:02}, \citealt{belczynski:08}, and see also \citealt{geller:13} and \citealt{eggleton:06}).  
Assuming a uniform mass-ratio distribution, this critical mass ratio allows only 2/3 of the binaries to
potentially undergo stable mass transfer, and thereby provides a factor of $f_\text{q} = 2/3$ below.
These factors, multiplied by the rate at which stars evolve off the MS at the given cluster age and metallicity
  ($\Gamma_\text{ev}$, see Section~\ref{s:pchan}), yield
\begin{equation}\label{e:tau_MT}
\tau_\text{SG MT} = \left(\Gamma_\text{ev} f_\text{b} f_\text{P} f_\text{q}\right)^{-1} .
\end{equation}

\subsection{Increased Magnetic Activity in a Subgiant Star}
\label{s:pSGMag}

To calculate $\tau$, we follow a similar method as in Section~\ref{s:pSGMT} to estimate the mean time between stars in appropriate binaries 
evolving off of the MS.  
Here, for $f_{\rm P}$, we set the short-period limit to be that at the Roche radius (see Equation~\ref{e:roche}, thereby excluding any binaries
included in Section~\ref{s:pSGMT}) and the
long-period limit to the binary circularization period of the cluster.  
We estimate the circularization period of a cluster of 
a given age from the results of \citet[][dotted line in their Figure 2, that matches the observed binary circularization periods from 
\citealt{meibom:05} ]{geller:13}.  
The fraction of binaries with these short periods defines $f_\text{P}$.  We allow all mass ratios here.

However, not all short-period binaries containing a subgiant star must become SSGs.  A sample of the open clusters 
(NGC 188, NGC 2682, NGC 6819, and NGC 6791) have sufficient time-series radial-velocity and/or photometric observations to 
count the known binaries with orbital periods less than 15 days amongst the SSGs and subgiants, as a rough estimate of the efficiency of SSG formation 
through this mechanism. Within these clusters, we find four normal subgiants and nine SSGs, respectively, in binaries with periods $<$15 days.  We 
apply this fraction of $\alpha = 9/13$ to our calculation:
\begin{equation}\label{e:tau_mag}
\tau_\text{SG Mag} = \left(\alpha \Gamma_\text{ev} f_\text{b} f_\text{P}  \right)^{-1} .
\end{equation}

Finally, as noted above, here we again simply take $t$ as the lifetime of $S_1$ on the subgiant branch.  It is possible that such stars can 
remain in the SSG region also during the early evolution of the red-giant phase.  Adding this to $t$ would increase our probabilties of observing a SSG from ``SG Mag''.

\begin{figure*}[!t]
\includegraphics[width=7in]{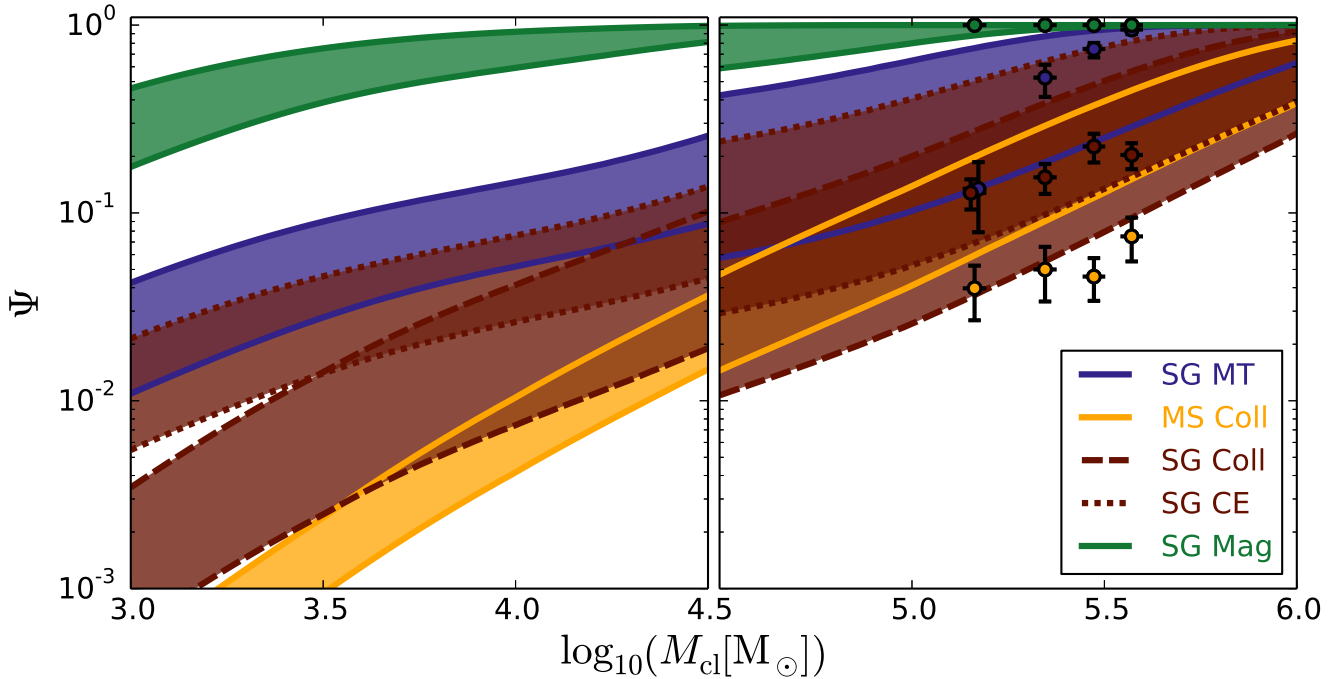}
\caption{
Poisson probabilities of observing SSGs resulting from the formation channels discussed in Section~\ref{s:fchan}.
``SG MT'' (blue) is the probability of observing a binary in the process of mass transfer from a subgiant donor (Sections~\ref{s:SGMT} and~\ref{s:pSGMT}). 
``MS Coll'' (yellow) is the probability of observing a MS-MS collision product before it settles back to the ZAMS (Sections~\ref{s:MSColl} and~\ref{s:pMSColl}).
``SG Coll'' and ``SG CE'' (red) are the probabilities of observing a subgiant after having its envelope rapidly stripped (Sections~\ref{s:SGStrip} and~\ref{s:pSGStrip}),
either through a grazing collision (dashed) or a common-envelope ejection (dotted). 
``SG Mag'' (green) is the probability of observing a subgiant with a reduced convective mixing length from enhanced magnetic activity (Sections~\ref{s:SGMag} and~\ref{s:pSGMag}).
Each region shows the Poisson probabilities derived from the weighted average timescales ($t$ and $\tau$ from Equation~\ref{e:poisson}) over our grid of models, 
weighted by the observed distributions of ages, half-mass radii and metallicities for open clusters (left) and 
globular clusters (right), as described in Sections~\ref{s:pchan} and~\ref{s:pgen}.  The widths show one (weighted) standard deviation above and below the weighted mean. 
Additionally we plot predictions from globular cluster Monte Carlo models for the probability of observing SSGs created through each channel 
(see Section~\ref{s:nbody}): points show the weighted means, vertical error bars show the standard errors of the mean, and horizontal bars show the widths of each mass bin.
(Mass bins are the same for each channel; for the lowest-mass bin of the ``SG MT'' and ``SG Coll'' channels, we shift the points slightly for readability.)
\label{f:poisson}
}
\end{figure*}

\subsection{Rapid Mass Loss from a Subgiant Star}
\label{s:pSGStrip}

Here we investigate two stripping mechanisms: through (i) common-envelope or (ii) a grazing collision.  
For the common-envelope case, ``SG CE'', we use nearly the same calculations as for the ``SG MT'' channel (Section~\ref{s:pSGMT}), but 
here we set $f_\text{q} = 1/3$ in Equation~\ref{e:tau_MT}.  This optimistic scenario assumes that every subgiant that undergoes a common-envelope will have its envelope stripped in such a way as 
to produce a SSG.  

For the grazing collision case, ``SG Coll'',
$\tau$ is the mean time between collisions involving the stars of interest:
\begin{equation}\label{e:tcol}
\tau_\text{SG Coll}\left(a\right) = \left[f_\text{SG} \left(2\Gamma_{11} + 3 f_\text{c12} \Gamma_{12}\left(a\right) + 4 f_\text{c22} \Gamma_{22}\left(a\right)\right)\right]^{-1} ,
\end{equation}
where $\Gamma_{11}$, $\Gamma_{12}$, and $\Gamma_{22}$ are the single-single, single-binary, and binary-binary encounter rates from
\citet{leigh:11}\footnote{The encounter rates depend on the binary fraction, core radius, stellar density, velocity dispersion, mean stellar mass, and the
physical size of the object (i.e., the stellar radius for a 1+1 encounter and the semi-major axis for the 1+2 or 2+2 encounter).  We describe how we estimate these values
in Section~\ref{s:pchan}.},
(and $\tau = 1/\Gamma$), except here we multiply each rate by a factor, $\left(N f_\text{SG}\right)$, to account for the requirement that at least one of the stars involved 
must be a subgiant, where $N$ = 2,3,4 is the number of stars in the encounter, and  $f_\text{SG}$ is the fraction of stars in the 
cluster that are expected to be subgiants (as explained above). 
$f_\text{c12}$ and $f_\text{c22}$ are the fractions of 1+2 and 2+2 encounters, respectively, that result in direct collisions, taken 
from the grid of scattering experiments of \citet{geller:15a} for a given cluster mass and half-mass radius.  As these scattering experiments only 
include MS stars, we multiply these factors by the ratio of the gravitionally-focused cross section for $S_1$ to that of a MS star at the turnoff.
\citep[i.e., $\left(M_{\rm S_1}R_{\rm S_1}\right)/\left(M_{\rm MSTO}R_{\rm MSTO}\right)$][]{leonard:89}.

$\Gamma_{12}$ and $\Gamma_{22}$ both depend on the binary semi-major axis, $a$, (or orbital period), and we allow 
binaries from the Roche limit of $S_1$ on the ZAMS up to the hard-soft boundary (thereby excluding encounters with soft binaries).  
To calculate $\tau_\text{SG}$ for Equation~\ref{e:poisson}, we take the average of 
$\tau_\text{SG Coll}\left(a\right)$, weighted by the log-normal period distribution (within the appropriate Roche limit and hard-soft boundary).  

We assume here that each collision results in sufficient stripping to produce a SSG.  This is likely an overestimate of the true SSG production
rate through this mechanism.  Again, we aim for the most optimistic assumptions in our calculations here.

Finally, as mentioned above we set $t_{\rm SG\ Strip}$ equal to the lifetime of $S_1$ on the subgiant branch.
In our exploratory MESA modeling in Paper II, we see that for different amounts of stripping, and for different assumptions about 
the time the stripping occurs, the product can have a lifetime in the SSG region that is somewhat greater than, or less than, the subgiant lifetime of $S_1$.  
Accounting for this level of detail is beyond the scope of this paper, but may warrant future investigation.

\subsection{Collision of two MS Stars}
\label{s:pMSColl}

To estimate $t$ here, we start with the mean time of all collision products in \citet{sills:97} to evolve from immediately after the collision back to the 
ZAMS, $t_{c0} = 6.74$ Myr.  
The mean increase in luminosity for all collision products in \citet{sills:97} from immediately after the collision until settling back to the MS is a factor of 10$^{1.5}$,
and we assume this increase for all collision products in our calculations.
We then make the simplifying assumption that the product's luminosity decreases linearly in time.
Finally we step through bins in stellar mass and calculate a weighted average of the time that a MS-MS collision product is estimated to 
remain in the SSG region for a given cluster:
\begin{equation} \label{e:tcoll}
t_\text{MS Coll}  = \frac{\sum_{m=m_0}^{m_\text{f}} \left(\frac{6.74}{\text{[Myr]}}\right) f(m) w(m)}{\sum_{m=m_0}^{m_\text{f}} w(m)} ,
\end{equation} 
where $w(m)$ weights by the mass function at the mass $m$,
$m_\text{f}$ is the ZAMS mass of $S_1$ and $m_0$ is the mass of a MS star with a luminosity that is 10$^{1.5}$ times smaller than $m_\text{f}$ (from SSE).
The factor $f(m)$ is an estimate of the fraction of the time from collision to ZAMS that the product is expected to remain in the SSG region;
this factor follows from our assumption that the luminosity of the product immediately after the collision increases by a factor of 10$^{1.5}$ then
decreases back to the ZAMS linearly with time, and may pass through the SSG region that extends from the magnitude of the main-sequence turnoff down to
1.5 magnitudes fainter (approximately covering the region of observed SSGs, see Figure~\ref{f:CMDreg}).  Certainly a more detailed treatment of this factor
is desirable, but is beyond the scope of this paper.

We follow the same approach to calculate $\tau_\text{MS Coll}$ as in Section~\ref{s:SGStrip}, but take $f_\text{c12}$ and $f_\text{c22}$ directly from \citet{geller:15a}, 
and use the fraction of MS stars with masses between $m_0$ and $m_\text{f}$, in place of the fraction of subgiant stars ($f_\text{SG}$), in the cluster.

\section{Comparison of the Probabilities of Observing Each Product}
\label{s:prob}

We use two methods to compare the probabilities of observing at least one product of each respective formation channel (given 
the two timescales for each channel discussed above):
one general and averaged over all observed open and globular clusters as a function of cluster mass (Section~\ref{s:pgen} and Figure~\ref{f:poisson}) 
and the other specific to each cluster with observed SSGs (Section~\ref{s:pclu}, Table~\ref{probtab} and Figure~\ref{f:bar}).

\subsection{General}
\label{s:pgen}

We begin by producing a grid of timescales ($t$ and $\tau$ from Equation~\ref{e:poisson}),
for each mechanism covering the range of relevant cluster ages (from 2 to 13 Gyr, in steps of 1 Gyr), 
masses (from $\log(M_\text{cl}$ [\Msolar]$) = $~3 to 6, in steps of 0.01), half-mass radii (from $r_\text{hm} = $~1 to 10 pc in steps of 1 pc)  
and metallicities (from [Fe/H] = -2.3 to 0.2, with steps of 0.5 for [Fe/H] between -2 and 0; the metallicity range possible in SSE is Z = [0.0001, 0.03], which 
corresponds to [Fe/H] $\sim$[-2.3, 0.2]) for observed open and globular clusters.  
We use a Plummer model and EMACSS, where necessary, and the assumptions discussed in Section~\ref{s:fchan}.

We then compile all available observed values of age, $r_\text{hm}$ and [Fe/H] for open \citep{salaris:04,vandenbergh:06}\footnote{\footnotesize We
note that a larger catalog for these parameters exists in \citet{piskunov:08} and \citet{kharchenko:13}, but here we are more interested in the older open
clusters, like those observed to have SSGs, which were more carefully analyzed and provided in the given references.} and globular \citep{marin:09,harris:96,harris:10} clusters.  
Then for each of these two samples, we take a weighted average of our 
calculated grid of timescales for each respective mechanism, weighted by the fraction of open or globular clusters within each bin of age, $r_\text{hm}$ and [Fe/H].
Finally, we use these weighted average timescales to calculate the Poisson probabilities of observing at least one SSG within a cluster of the given mass.
We divide our results at a mass of $10^4$ \Msolar, which separates our sample at roughly the transition mass between open and globular cluster mass.  

The resulting probabilities for each SSG formation mechanism are shown in Figure~\ref{f:poisson}, in the different colored regions with widths equal to 
one (weighted) standard deviation from the weighted mean value. 
In general, taking 1/$\Psi$ gives the number of clusters that should be observed in order to expect 
to detect at least one SSG from the given mechanism.  Our calculations predict that roughly one in every few 
open clusters and nearly every globular cluster should host at least one SSG. This is in reasonable agreement with the current state of observations 
(see Figure~\ref{f:NvM} and Section~\ref{s:pclu}), 
though no systematic survey for SSGs exists (in open or globular clusters).  As we've taken optimistic assumptions in our calculations, these probabilities may be
interpreted as upper limits.

Our calculations predict that the probability of observing SSGs from all mechanisms will increase with increasing cluster mass.  This is simply due to the larger number of 
stars.  More importantly, for clusters of all masses, we predict that isolated binary evolution mechanisms are dominant.   
The other mechanisms follow at lower probabilities; though toward the highest-mass globular clusters, it becomes equally likely to observe at least one SSG from 
all mechanisms.  

Although we show in Figure~\ref{f:poisson} the probabilities of observing SSGs as a function of cluster mass, cluster density (and encounter rate) is also important. 
For a given cluster mass, the rate of SSG formation through the collision channels increases with increasing density, while the 
rate of SSG formation through the binary evolution mechanisms is nearly independent of density (within the range of 
parameters relevant to observed open and globular clusters).  
The only dynamical mechanism that can affect the binary evolution channels in these calculations 
is the truncation of the binary orbital period distribution at the hard-soft boundary, which, for clusters of interest, 
is at longer periods than the synchronization period 
(and the period at Roche lobe overflow).  Again, these are optimistic assumptions meant to provide an upper limit on SSG formation rates.
As we discuss below, more subtle dynamical effects, like perturbations 
and exchanges, within hard binaries may decrease the true SSG production rate through the binary evolution channels for the most massive clusters.

\subsection{Cluster specific}
\label{s:pclu}

\begin{figure}[!t]
\includegraphics[width=3.5in]{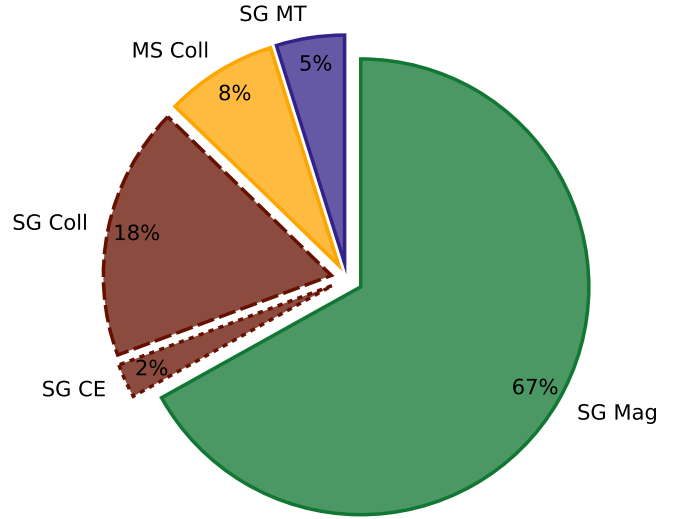}
\caption{
Percent of total SSGs predicted from each formation mechanism (see Sections~\ref{s:fchan} and~\ref{s:pchan})
in all
the observed clusters in Table~\ref{probtab}.
\label{f:bar}
}
\end{figure}

\begin{figure}[!t]
\includegraphics[width=3.5in]{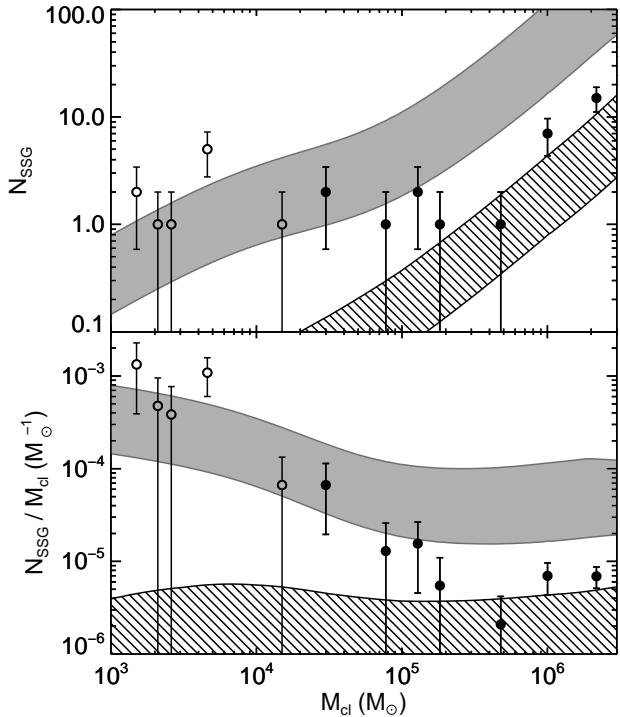}
\caption{
Number (top) and specific frequency (bottom; number of SSGs, $N_\text{SSG}$, divided by the cluster mass, $M_\text{cl}$) of SSGs as a function of the cluster mass.
Observed open/globular clusters from Paper I are plotted in open/filled symbols.
As in Paper I, we show only those observed SSGs with the highest-likelihood of cluster membership and within the same radial
completeness limit of $<3.3$ core radii (see Paper I for details).  Error bars show the standard Poisson uncertainties on $N_\text{SSG}$
(and we truncate the lower error bars for cases with $N_\text{SSG} = 1$).
The gray-filled region shows the predicted number of SSGs from our calculations in Section~\ref{s:pgen} through all mechanisms combined, and the 
hatched region shows the predicted number of SSGs for the collision mechanisms alone (i.e., ``SG Coll'' and ``MS Coll'').
Note that our Poisson calculations are not limited in radius from the center of the cluster (as are the observations), and rely on 
optimistic assumptions; these calculations show upper limits.
\label{f:NvM}
}
\end{figure}

In addition to the general calculation described above, we also perform specific calculations of the respective probabilities to observe at least one 
product of each of the formation channels for each cluster with a SSG candidate.  Here we compile all available data for each cluster that would serve as an 
input into our probability calculations described in Section~\ref{s:pchan}, and provide these in Table~\ref{probtab}.  As described above, our calculations 
require the age, mass, metallicity and either the core or half-mass radius.  Where available, we provide the additional empirical input to our calculations 
of the observed binary frequency ($f_\text{b}$), central density ($\rho_0$), core radius ($r_\text{c}$), half-mass radius ($r_\text{hm}$) and circularization period ($P_\text{circ}$).
All other necessary values that are unavailable in the literature are inferred using the same assumptions as above.  

We use these empirical values to determine $t$ and $\tau$ in Equation~\ref{e:poisson}, as described in Section~\ref{s:pchan}, and provide the probabilities of observing 
at least one SSG from the given mechanism in each cluster in Table~\ref{probtab}.
We also provide the combined Poisson probabilities of observing the observed number of SSGs ($n_\text{SSG}$) in 
each cluster\footnote{\footnotesize The number of SSGs is taken from Paper I, where we select stars that reside in the SSG region of the CMD in at least one available 
color-magnitude combination and have a $<10$\% probability of being a field star.}
from any formation channel (calculated by summing the $t/\tau$ values from each mechanism and using this in Equation~\ref{e:poisson}, and only given for clusters with SSGs).
For ease of reading, we do not include uncertainties on these probabilities in the table; however we do follow the uncertainties on each input parameter through our 
calculations for each probability. If a parameter does not have uncertainties in the literature (and therefore no error is given in the table), we assume a 10\% uncertainty 
for our calculations.
The number of digits provided in the Table shows the order of magnitude of the inferred range in probabilities resulting from the uncertainties in input values.  
We round any probability $>0.99$ up to 1.

In Paper II, we investigate the SSGs in two of these clusters, NGC 6791 and M67, in depth and perform more careful calculations of their formation (involving more detailed empirical 
input and using a slightly different method).  
Our results here agree very well with those from Paper II, which provides further confidence in our calculations here.
Specifically, in Paper II we find a probability of observing at least one SSG from the ``SG MT'' mechanism in M67 of 4\% and in NGC 6791 of 14\%, where here we find 5\% and 9\%, 
respectively.  In Paper II, we find a probability of 42\% and 94\% of observing at least one SSG from the ``SG Mag'' in M67 and NGC 6791, respectively, as compared to 47\% and 82\% here.  
For the ``SG Coll'' scenario, in Paper II we find a probability of $\sim$3\% that we would observe at least one in M67, compared to 20\% here.  Though this particular probability value appears higher 
here (due to our more optimistic assumptions), the uncertainty on this probability is of the same order as the value itself.

Additionally, we show the results graphically in Figure~\ref{f:bar}, where we plot the percentage of SSGs predicted, over all clusters in Table~\ref{probtab}, to come from each mechanism.
To construct this plot, we sum the number of predicted SSGs for a given mechanism over the observed clusters, and divide by the total number of SSGs predicted for all 
clusters from all mechanisms.  For instance, our calculations predict that 67\% of sub-subgiants in these observed clusters may come from the ``SG Mag'' mechanism.

Nonetheless, if we sum the probabilities for each mechanism given in Table~\ref{probtab}, we expect to observe at least one SSG from each mechanism when considering all clusters.  
For nearly all of the globular clusters our calculations suggest that these formation channels are 
sufficient to explain all observed SSGs (i.e., $\Psi_{\rm nSSG}\sim1$ for these clusters).  
In the open cluster regime, the number of SSGs predicted for clusters in this mass range is in rough agreement with the observations (Figure~\ref{f:NvM}), 
though the specific $\Psi_{\rm nSSG}$ values for the observed open clusters are below one in Table~\ref{probtab}.
This may indicate that we have overlooked viable formation channels in the open cluster regime or that we have underestimated values in our calculations
primarily for open clusters, and we return to this in Section~\ref{s:discuss}.

In Figure~\ref{f:NvM} we show the number of SSGs predicted by our model as a function of cluster mass, compared to that of the observed clusters (see Paper I, Figure~7).
The gray band combines all formation channels, while the hatched region shows only the collision channels.  Our model agrees with the general trend in the observations,
of decreasing specific frequency of SSGs toward increasing cluster mass.  However toward the high-mass end, our model begins to over-predict the number of SSGs.  
This may imply that there are more SSGs to be discovered in these clusters (which indeed is expected, see Paper I).
This discrepancy may also be tied, at least in part, to our simplified treatment of how dynamics affects the binary evolution channels.  Perhaps more 
subtle dynamical effects (such as perturbations or exchanges, not included in these calculations) inhibit the binary evolution 
channels significantly in clusters with high encounter rates (like the massive observed clusters in our sample).
We investigate this further in the following section.
Indeed, for the most massive clusters in our sample, our model predicts that the collision mechanisms alone can nearly produce the observed numbers of SSGs.

\section{Sub-subgiants in Star Cluster $N$-body models} 
\label{s:nbody}

%

\begin{figure}[!ht]
\includegraphics[width=3.41in]{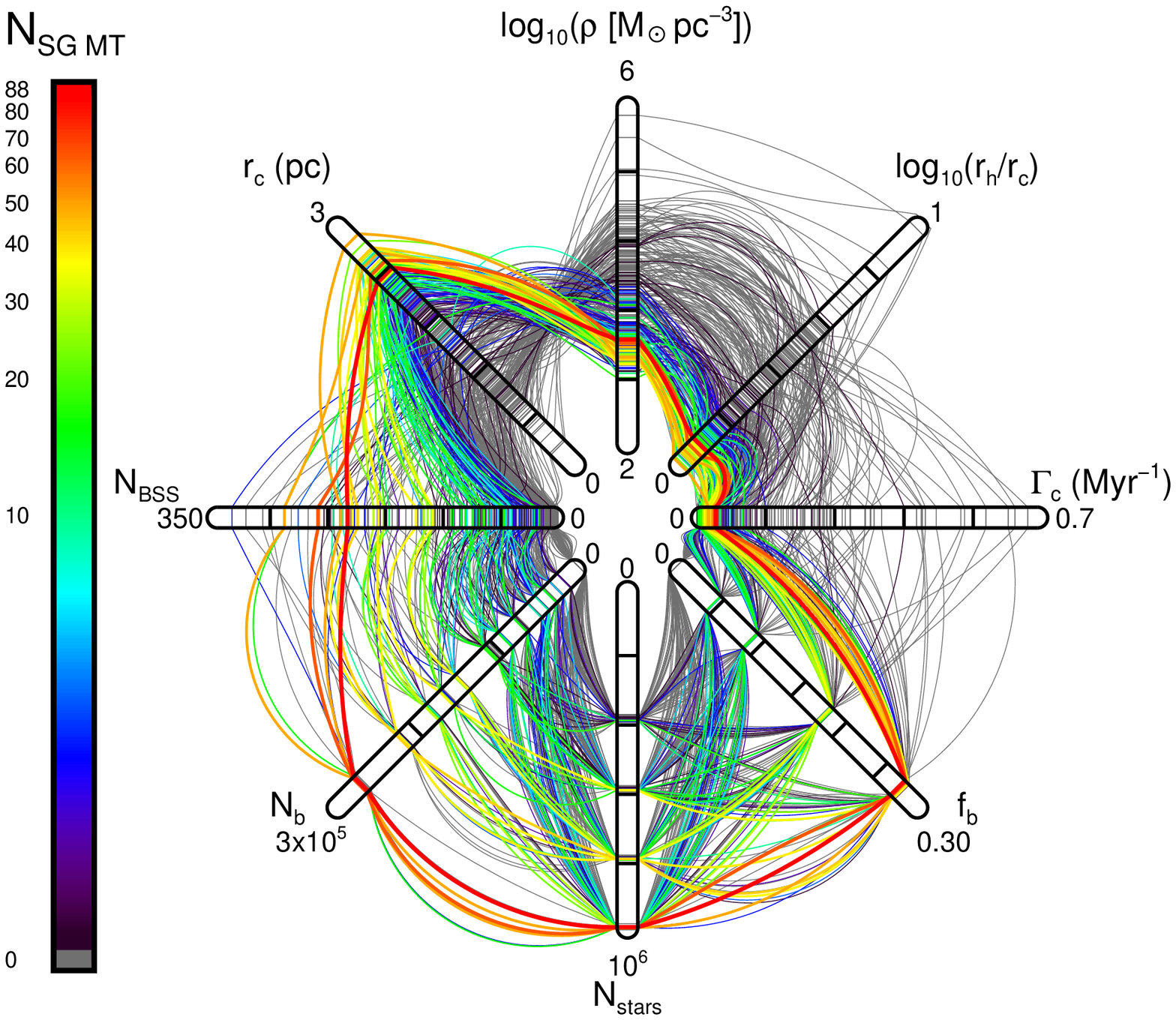}
\includegraphics[width=3.41in]{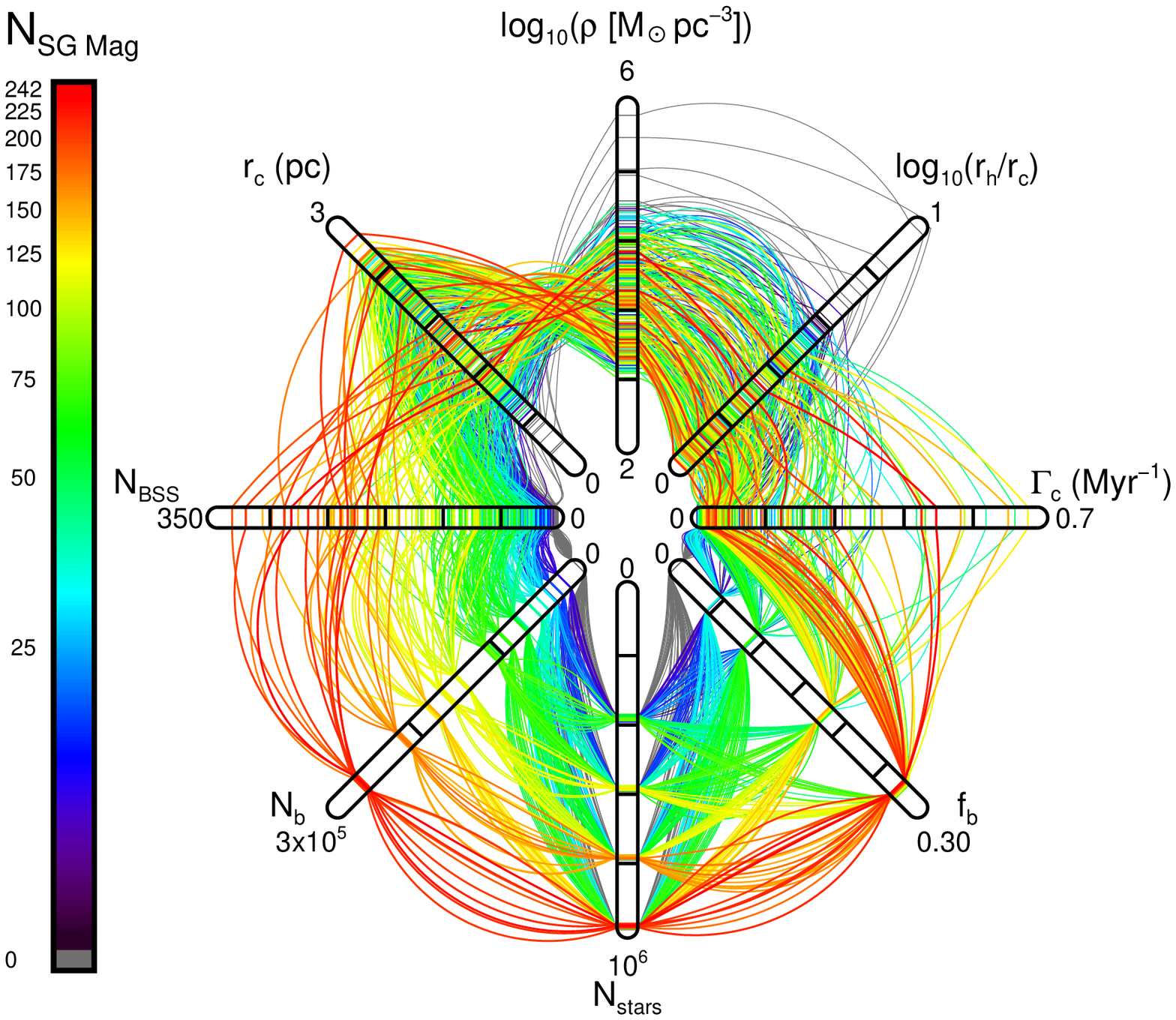}    
\caption{
Comparison of the number of SSGs from the ``SG MT'' ($N_\text{SG MT}$) and ``SG Mag'' ($N_\text{SG Mag}$) channels,
created in a grid of Monte Carlo globular cluster models that have the given
total numbers of stars ($N_{\rm stars}$),
binaries ($N_{\rm b}$) and 
blue straggler stars ($N_{\rm BSS}$),
core radius ($r_{\rm c}$),
central density ($\log_{10}(\rho)$),
ratio of the half-mass and core radii ($r_{\rm h}/r_{\rm c}$),
core collision rate ($\Gamma_{\rm c}$), and
core binary frequency ($f_{\rm b}$).
These parameters are all calculated theoretically
at the same snapshot times as we use to identify the SSGs,
and some may be slightly different from what an observer would measure \citep{chatterjee:13}.
We show network diagrams for each channel (top: ``SG MT'', bottom: ``SG Mag''), where each path around the plot defines a specific cluster model, crossing the
axes at the given cluster parameters, and colored by the number of SSGs created by that channel (see color bars at left of each plot).
\label{f:CMC1}
}
\end{figure}

\begin{figure*}[!t]
\begin{tabular}{lcr}
\includegraphics[width=2.3in]{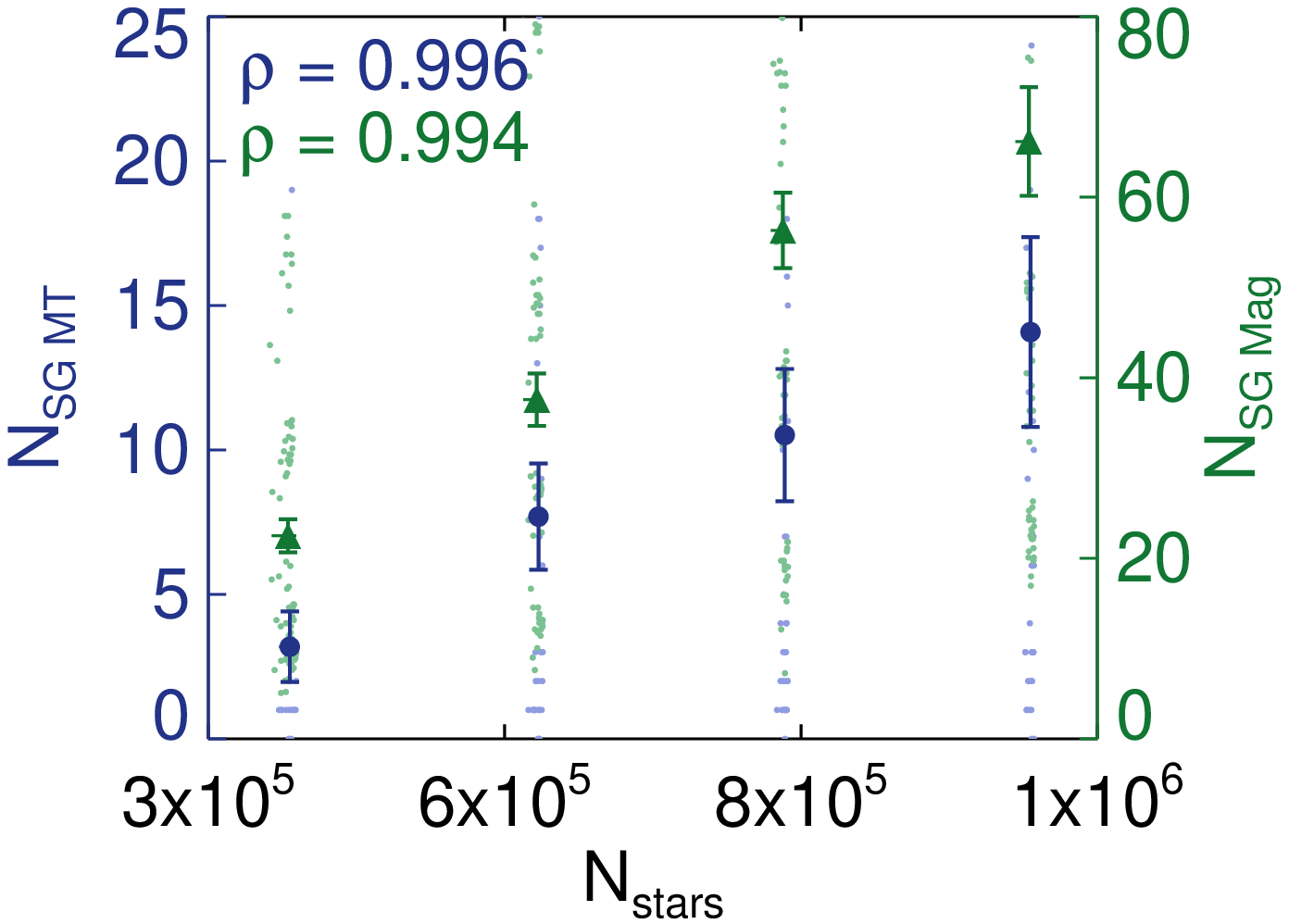} & \includegraphics[width=2.3in]{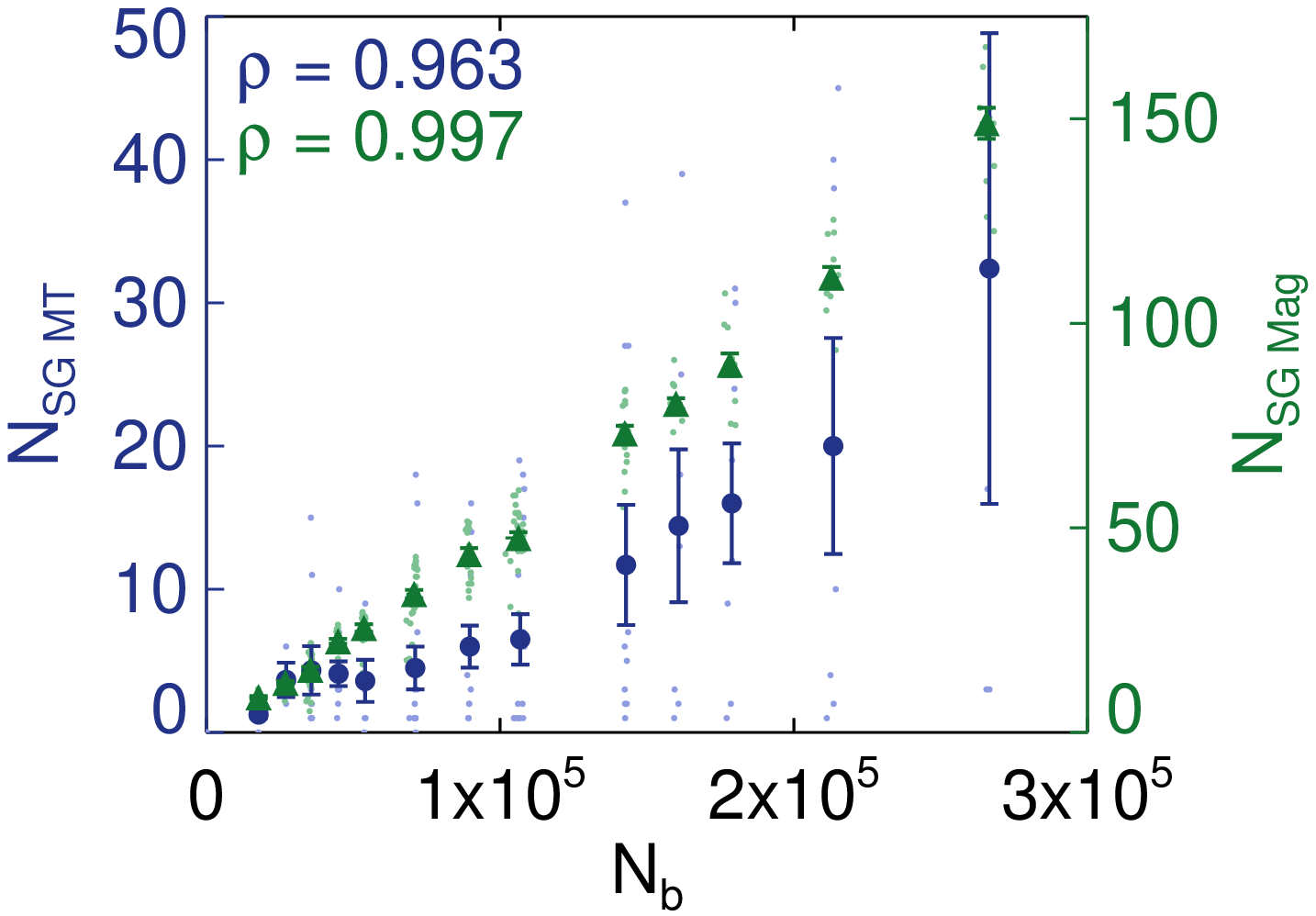} & \includegraphics[width=2.3in]{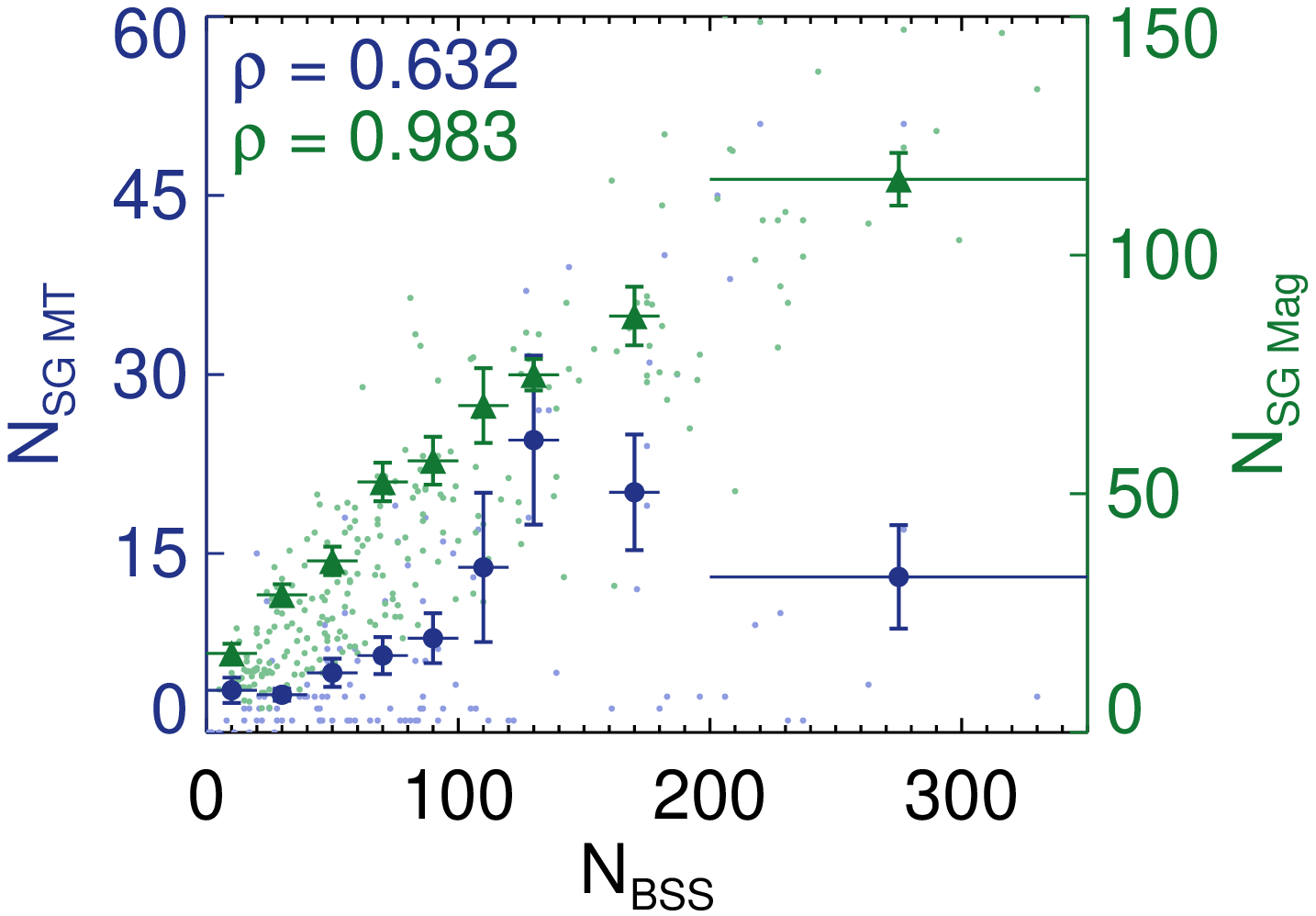} \\[0.5em]
\includegraphics[width=2.3in]{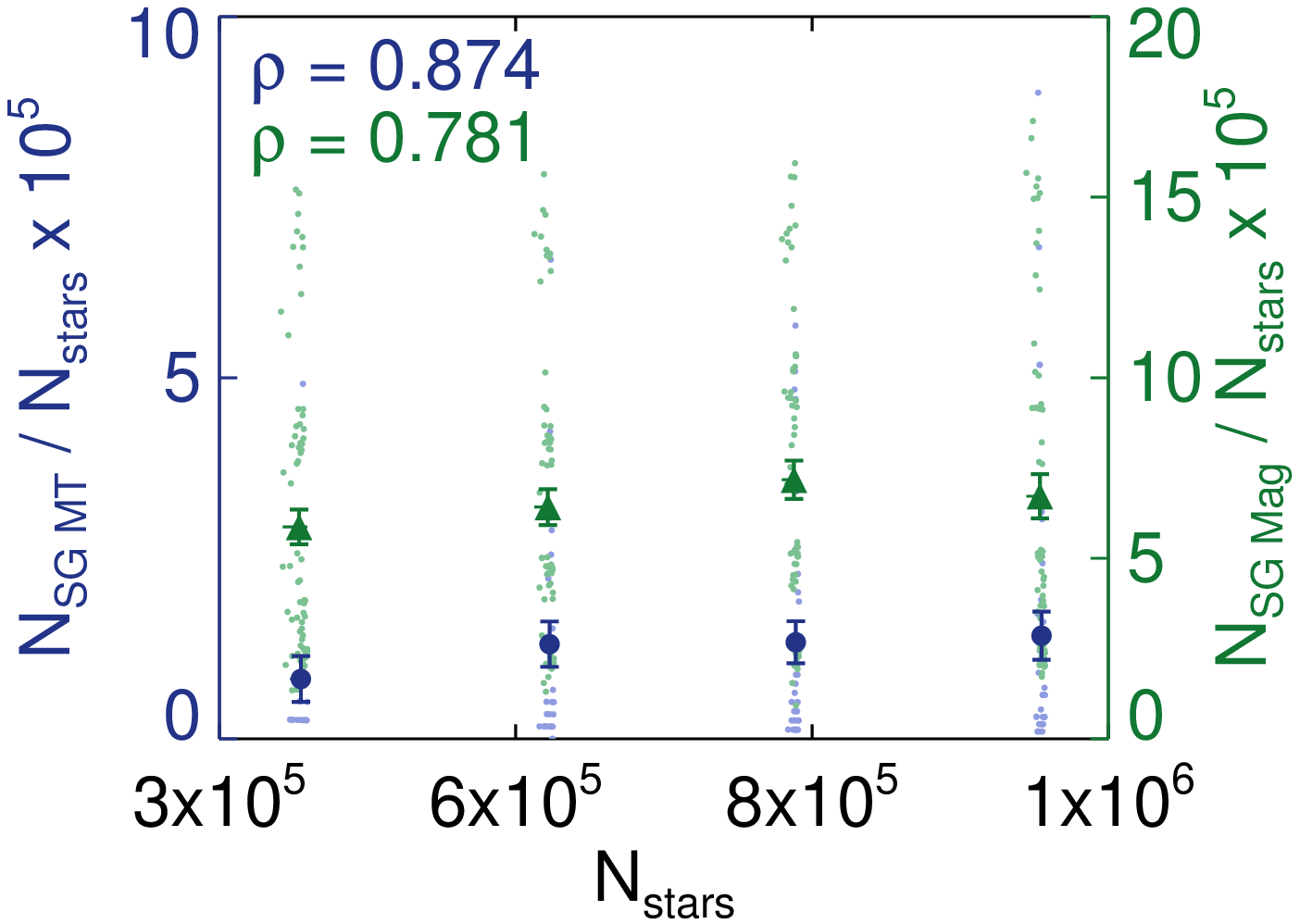} & \includegraphics[width=2.3in]{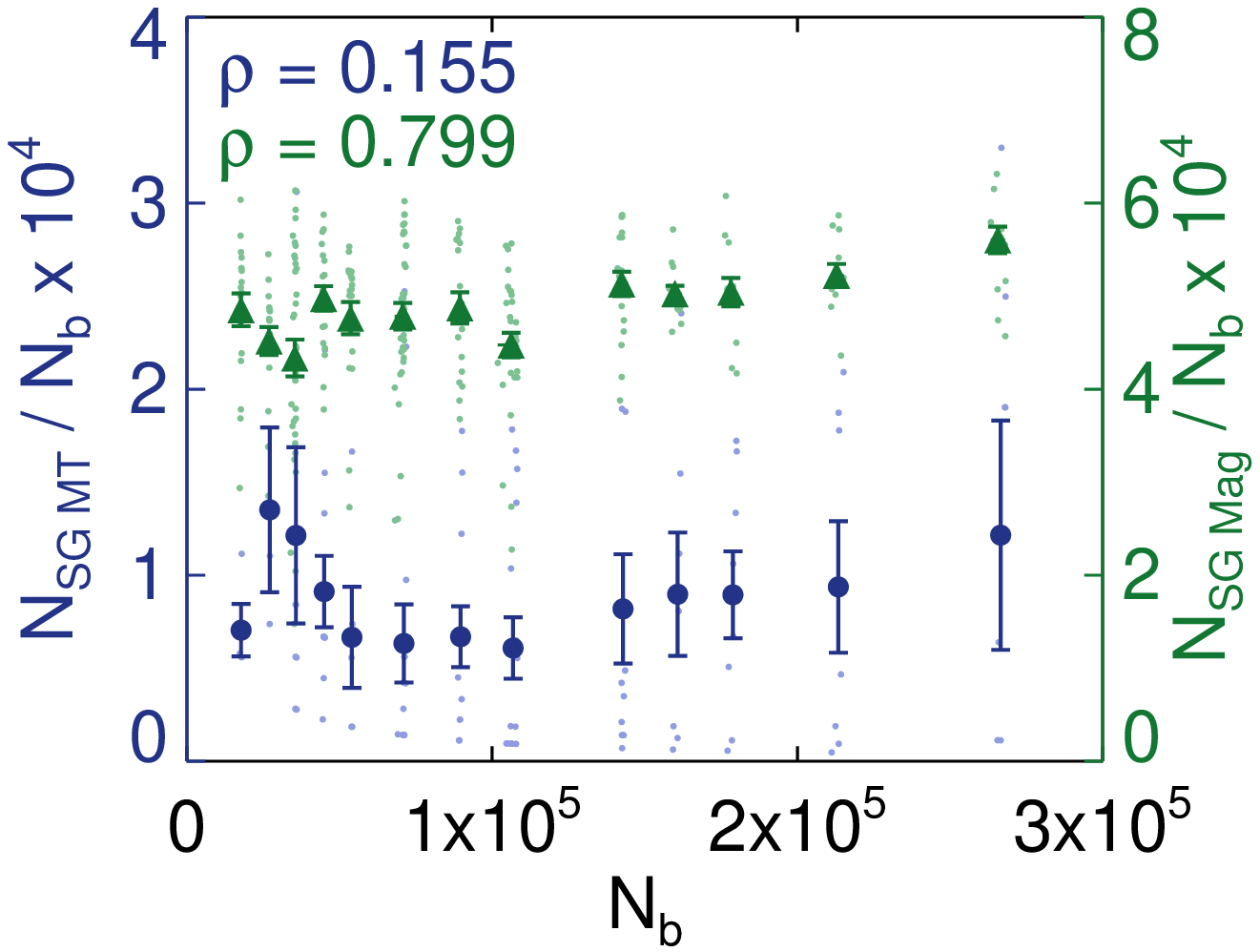} & \includegraphics[width=2.3in]{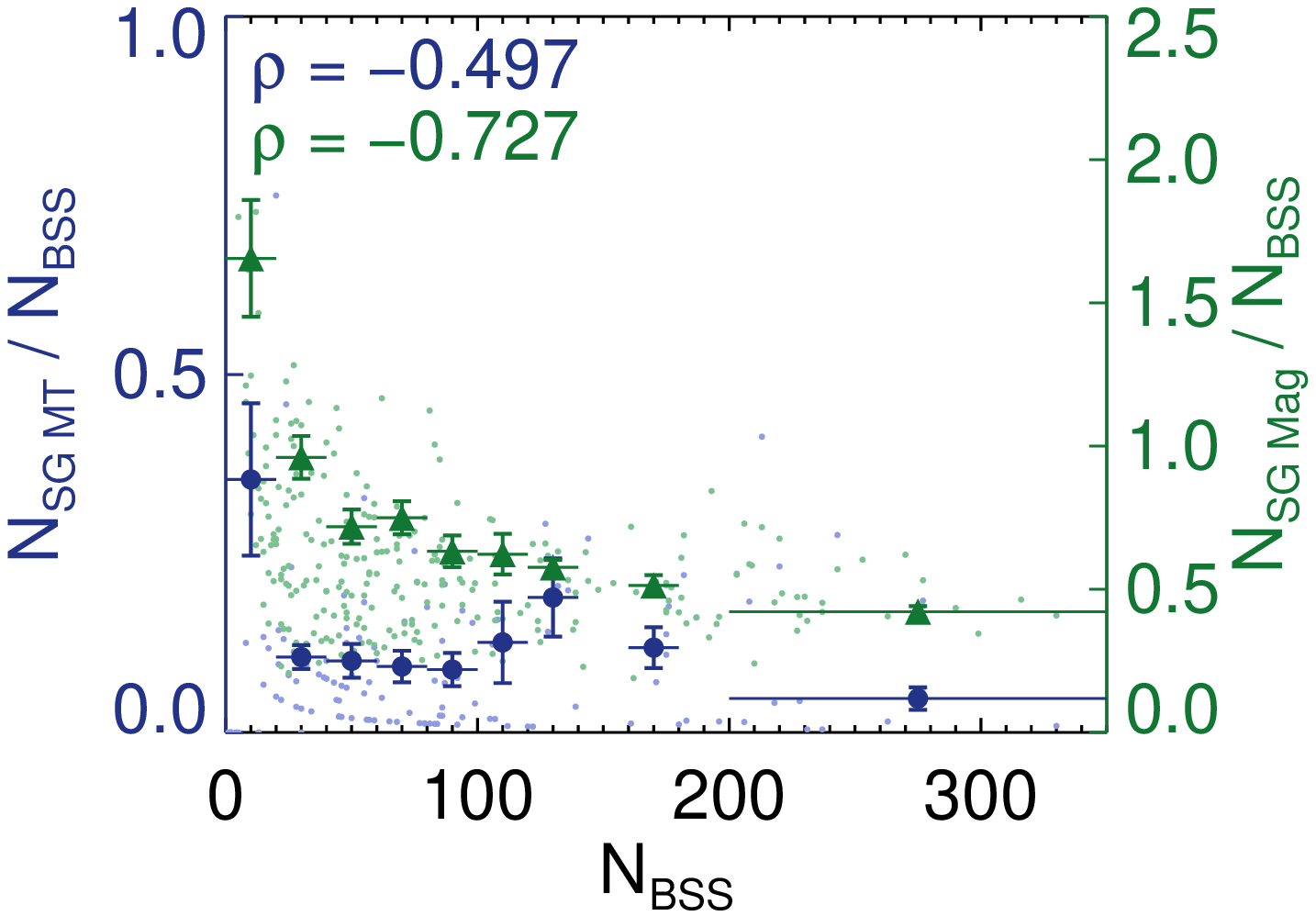} \\[0.5em]
\includegraphics[width=2.3in]{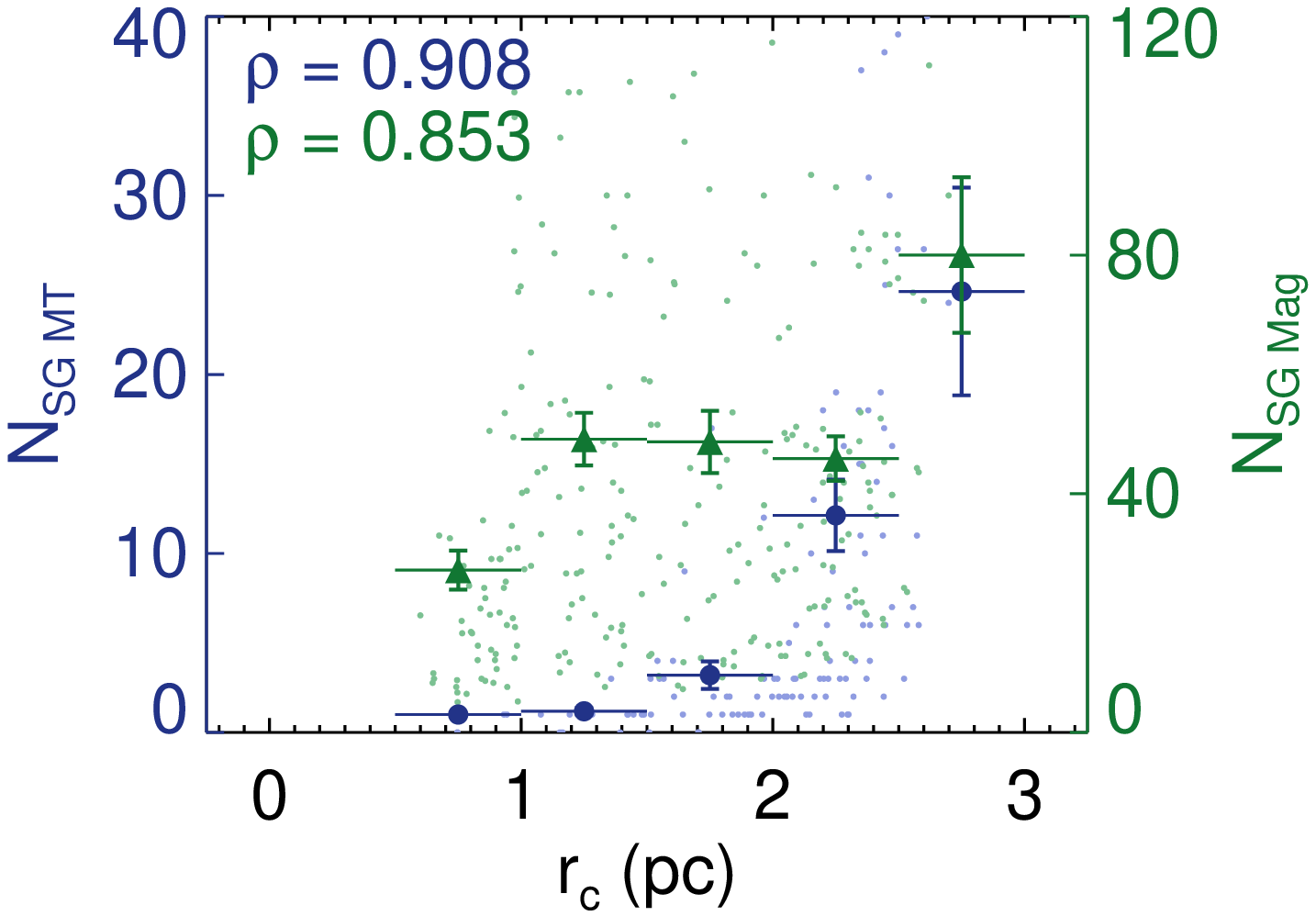} & \includegraphics[width=2.3in]{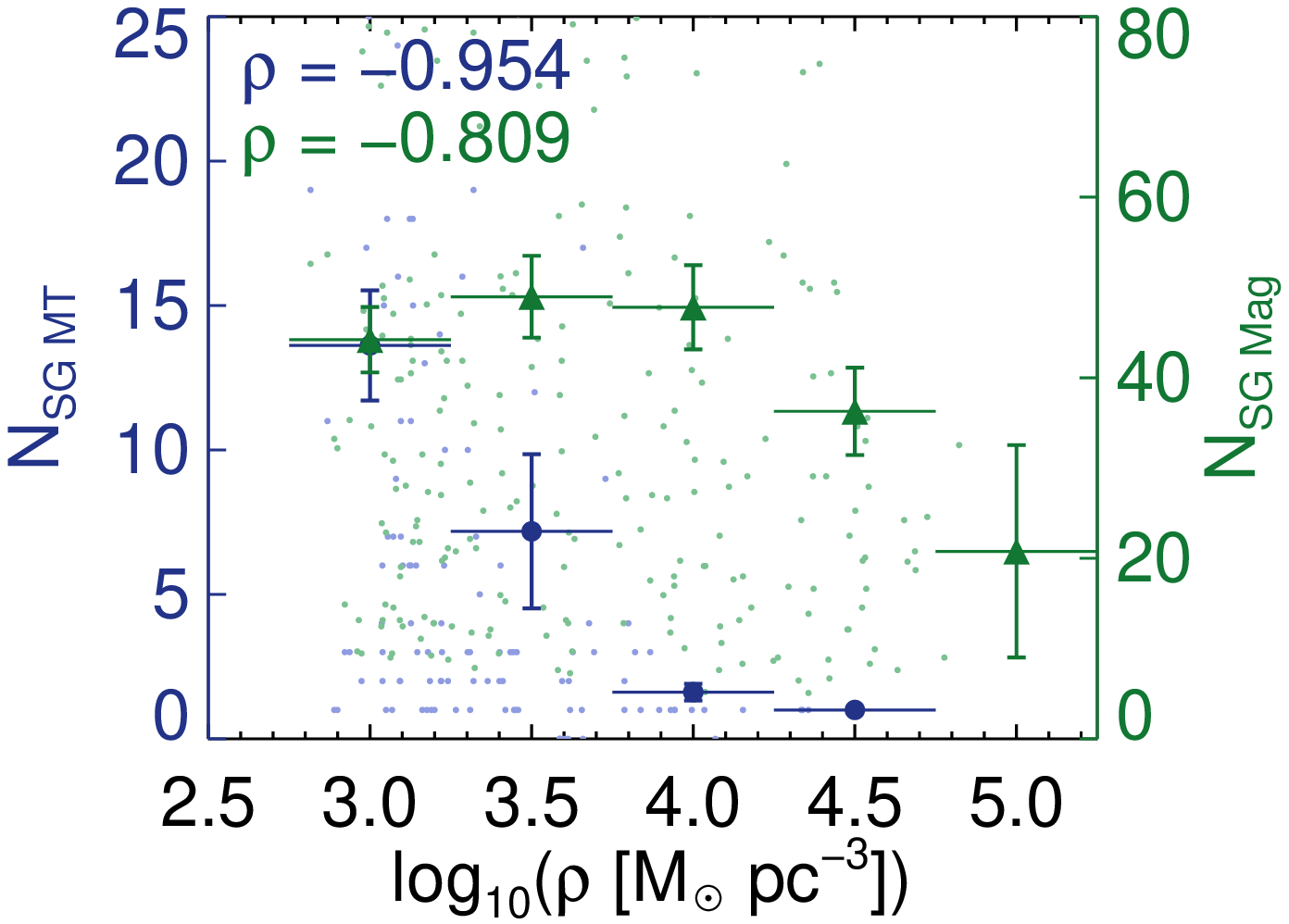} & \includegraphics[width=2.3in]{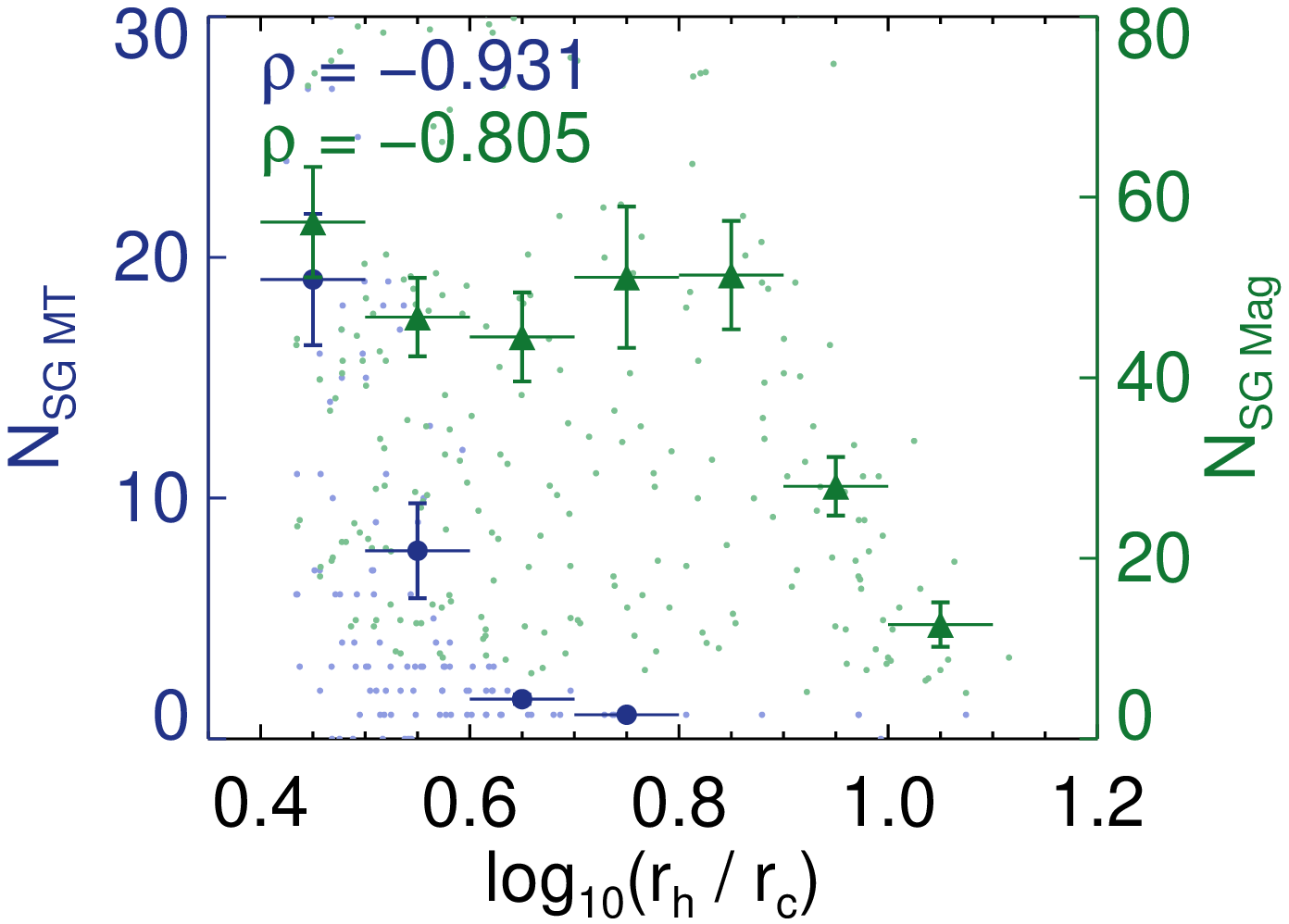} \\[0.5em]
\includegraphics[width=2.3in]{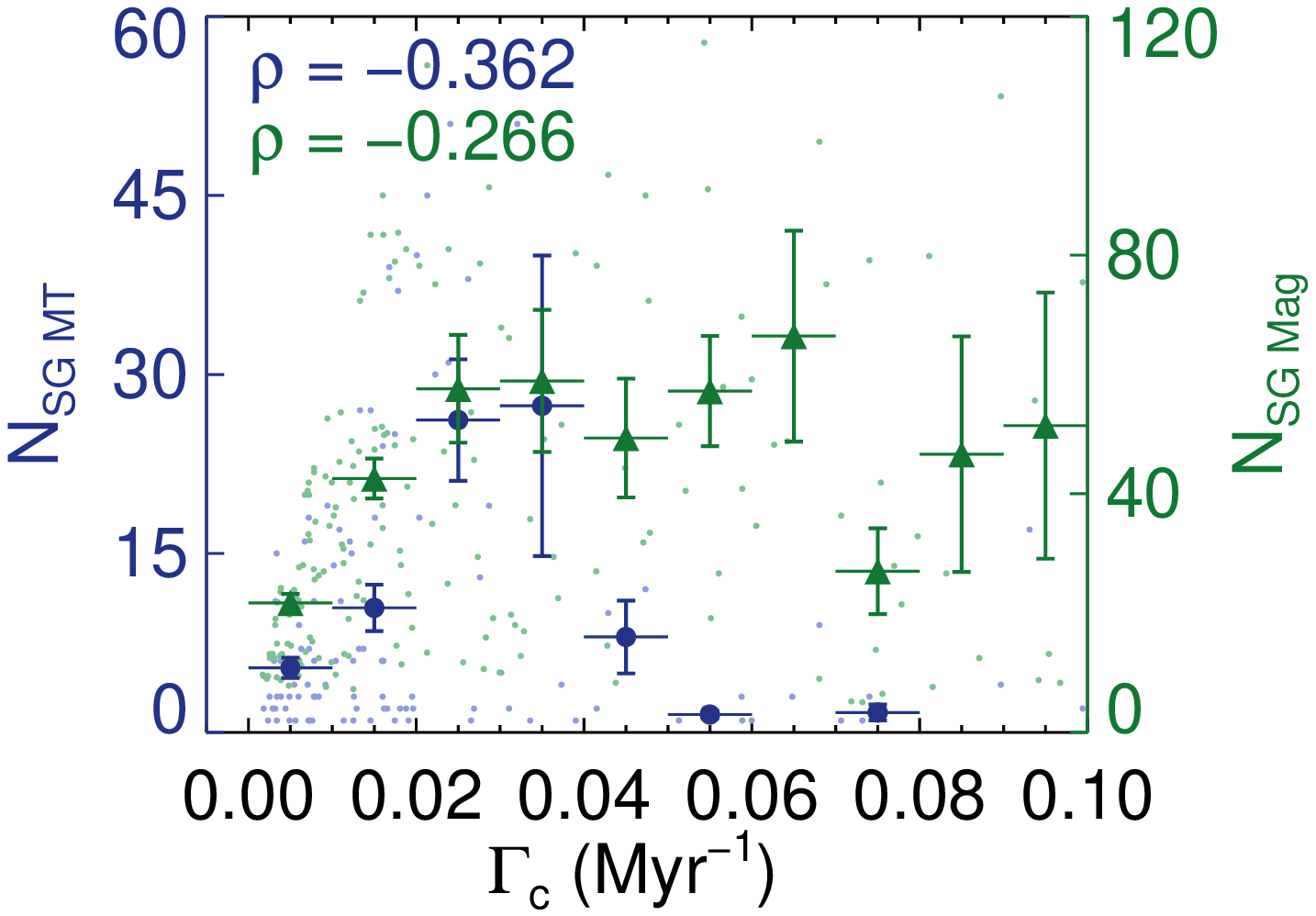} & \includegraphics[width=2.3in]{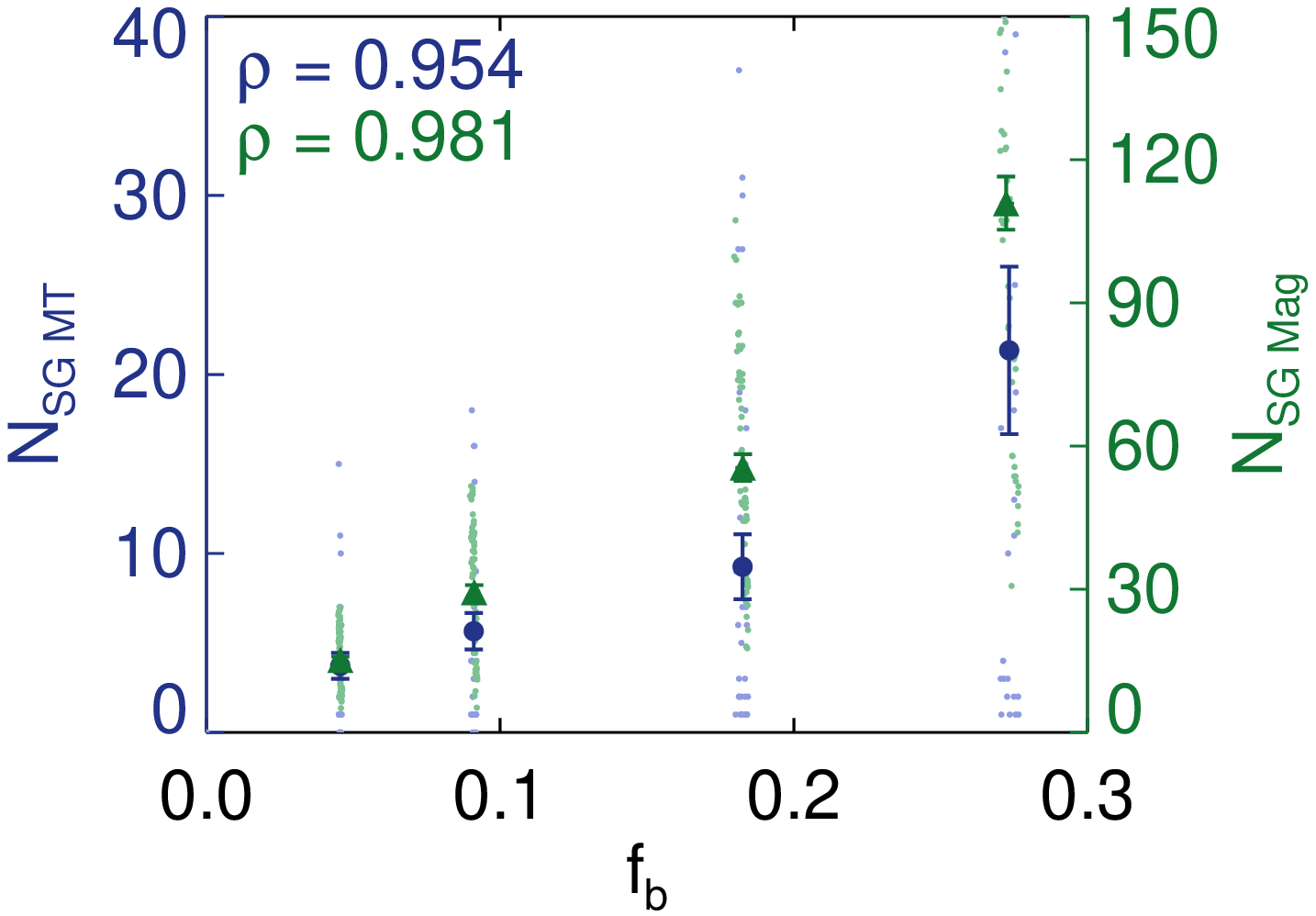} & \\[0.5em]
\end{tabular}
\caption{
Comparison of the number of SSGs from the ``SG MT'' ($N_\text{SG MT}$, blue circles) and ``SG Mag'' ($N_\text{SG Mag}$, green triangles) channels,
created in a grid of Monte Carlo globular cluster models and showing the same parameters as in Figure~\ref{f:CMC1}.
Here we plot the number of SSGs as a function of each of these parameters respectively, showing only models that produced at least one SSG.
Small points show the raw values from the grid, and larger points show the mean values in bins, with vertical error bars equal to the standard errors of the mean
and horizontal lines showing the bin sizes (which are smaller than the symbols in some cases).
For reference, we also include the respective Pearson correlation statistics ($\rho$), calculated for the mean values, in each panel.
}
\label{f:CMC2}
\end{figure*}

Our Poisson probability calculations make simplifying assumptions about SSG formation,
and provide upper limits for SSG formation rates.  $N$-body star cluster models can alleviate some of these simplifications, and
in particular can allow us to study the effects from more complex dynamical encounters and subtle perturbations, that we do not consider in our analytic calculations. 

\subsection{Direct $N$-body Models}
To our knowledge, the \citet{hurley:05} $N$-body model of M67 is the only star cluster model that specifically discusses the creation of a SSG star.
They used the \texttt{NBODY4} code \citep{aarseth:99}, which utilizes BSE \citep{hurley:02} for binary-star evolution. 
The only pathway available for SSG formation in these models is through binary evolution; the other mechanisms discussed here are not yet implemented in the $N$-body code
for SSG formation (though some are implemented to produce BSS).

This specific binary first went through a stage of conservative stable mass transfer, where the subgiant primary transferred mass 
onto its MS companion.  This then led to a common-envelope merger event that created the SSG single star seen at the age of M67. 
(We refer the reader to \citealt{hurley:05} for a more detailed description of this star's history.) 
This mechanism is similar, in part, to our ``SG MT'' pathway (Section~\ref{s:SGMT}), and is formally included in the ``SG MT'' rate calculations described in
Sections~\ref{s:pchan}~and~\ref{s:prob} (because the system starts with stable mass transfer).
Unlike our mechanism however, 
the \citet{hurley:05} star is more massive than the normal giants in the cluster at the age of M67, but with a lower core mass than the normal giants. 
They attribute the lower luminosity of the object to this lower core mass.   
Through our extensive BSE modeling (see Paper II), we do not see common-envelope merger products as a dominant SSG formation channel within the 
mass-transfer mechanism, though we have likely not covered the entire parameter space leading to SSG formation in BSE (and common-envelope evolution
remains poorly understood and only approximated within BSE).
Furthermore, as most of the observed SSGs in open clusters appear to be in short-period binaries, this specific 
pathway may not produce SSGs similar to the majority of those observed.

\subsection{Monte Carlo Models}

We also investigated a grid of Monte Carlo globular cluster models, from the Northwestern group \citep{joshi:00,joshi:01,fregeau:03,fregeau:07,chatterjee:10,umbreit:12}.
Specifically, we use a superset of the simulations presented in \citet{chatterjee:10,chatterjee:13a,chatterjee:13}, which includes 327 models
that cover the parameter space of the observed globular clusters in our Galaxy (though all at a metallicity of Z=0.001).
We examine snapshots from these models between 9 and 12 Gyr.
We used two methods to identify SSGs in these models: (i) we selected SSGs based on the location in the H-R diagram (as in Figure~\ref{f:CMDreg}),
and (ii) we identified other stars that may be observed as SSGs in a real cluster, but were not found in the SSG region of the simulated H-R diagram
due to limitations of BSE (which is used in both the \texttt{NBODY4} and Monte Carlo models).

Method (i) discovers all SSGs produced through the ``SG MT'' channel; this is the only mechanism available to producing SSGs within BSE.
We identified over 1100 ``SG MT'' SSGs in these models.  99\% of these simulated SSGs are currently in binaries, and the remainder were previously
in binaries.  98\% of the SSGs in binaries are currently undergoing RLOF.  Of the few that are detached, $\sim$80\% contain an evolved star that had previously 
lost $\geq$0.1\Msolar, presumably from a recently completed period of mass transfer (a subpopulation that we also briefly discuss in Paper II).
Importantly, only $\sim$10\% of these SSGs suffered strong encounters or direct collisions prior to becoming a SSG (though weak fly-bys are not
tracked in these models, as this is part of the relaxation process).  The vast majority of ``SG MT'' SSGs in these models avoided strong encounters
for the entire lifetime of the globular cluster. 

To investigate predictions for the other formation channels, we follow similar assumptions as in Section~\ref{s:pchan}.  More specifically,
we identify ``SG Mag'' SSGs as binaries in the models with orbital periods $P < P_\text{circ}$ that contain a subgiant (and then multiply
the number identified by our empirical fraction of $9/13$, see Section~\ref{s:pSGMag}).
We identify ``MS Coll'' SSGs as the products of collisions involving two main-sequence stars that occurred close enough in time to the model
snapshot output time and have a
product bright enough to reside in the SSG region (using the same assumptions as Section~\ref{s:pMSColl}).
Finally, we identify ``SG Coll'' SSGs as the products of collisions involving at least one subgiant star that occurred close enough in time
to the model snapshot output time.  (Likely not all of these collisions would create SSGs, but this will provide an upper limit.)
Through this method we identify more than
12000 additional SSGs,\footnote{\footnotesize Collisions are tracked continuously within these models, while full snapshot output occurs roughly
every Gyr; common-envelope events are not tracked continuously, and therefore we cannot investigate ``SG CE'' here.} primarily from the ``SG Mag'' channel.

We plot the Poisson probabilities of observing at least one SSG from these models in bins of cluster mass within Figure~\ref{f:poisson}.  
For the ``SG MT'' and ``SG Mag'' points, we first apply a correction factor to the number of SSGs in each model to account for a different
assumed binary orbital period (or semi-major axis) distribution; 
we assume a log-normal period distribution in Section~\ref{s:pchan}, while the Monte Carlo models use a distribution that is flat in the log.  For a given 
binary frequency, a flat distribution creates a factor of about 2.5 more short-period binaries (e.g., that can undergo RLOF on the subgiant branch) than does the 
log-normal distribution.  For all channels, we then take the average number of SSGs in each mass bin, 
weighted by the observed distributions of half-mass radii and cluster age (in a similar manner as described in Section~\ref{s:pgen}).  
We then set $t/\tau$ from Equation~\ref{e:poisson} equal to this weighted average number of SSGs from the models in each mass bin to calculate the Poisson probabilities.
The predictions from the Monte Carlo models agree well with those from our analytic upper limits from Figure~\ref{f:poisson},
even given the different assumptions that go into each method.
The Monte Carlo models predict a factor of a few less ``MS Coll'' SSGs than predicted analytically, likely due to our implicit assumptions in Section~\ref{s:pMSColl} of all encounters
occurring directly at the cluster center, and with zero impact parameter (neither of which are required in the Monte Carlo model).
Nonetheless, the agreement with this (relatively) independent method of deriving $\Psi$ for all channels supports the results of our more simplified analytic calculations.

As a further step, we also investigate the grid of Monte Carlo models for predictions of the type of clusters that should harbor the most SSGs.
The collision channels behave as expected, where more SSGs are produced in clusters with larger collision rates.  However, the vast majority of the
SSGs produced in all these Monte Carlo models ($>99$\%) derive from the binary evolution channels.
Furthermore, these models (plus our assumptions in identifying SSGs therein) predict on average about five times more ``SG Mag'' than ``SG MT'' SSGs.

We focus on these ``SG MT'' and ``SG Mag'' mechanisms here, and show detailed comparisons of these two channels in Figures~\ref{f:CMC1}~and~\ref{f:CMC2}.
Here we do not apply any correction to the number of SSGs from each model based on the input binary period distribution (as we did above).
Some of these Monte Carlo models contain very large numbers of SSGs, inconsistent with the (much smaller) number of SSGs observed
in the clusters we've studied.  This likely results from a combination of initial condition choices (some of which produce clusters that
don't match those we've studied), and also the details of binary evolution in BSE.  However, here we are not interested in the raw number of
SSGs produced; instead we investigate for trends in number of SSGs versus various cluster parameters predicted for these models.  

In Figure~\ref{f:CMC1}, we show network diagrams to 
visualize how all of the parameters from a given model relate to the number of SSGs created.  In this diagram, one arc around the figure corresponds 
to one model, hitting the axes at the appropriate values for the model, and with a color defined by the number of SSGs.
In Figure~\ref{f:CMC2}, we plot the number of SSGs against various (mostly observable) cluster parameters.

For both channels, we see correlations of increasing number of SSGs with increasing number of stars ($N_{\rm stars}$), number of binaries ($N_{\rm b}$)
and binary frequency ($f_{\rm b}$).
These correlations are expected, as nearly any population of stars that involve binaries (exotic or otherwise) should behave this way.
Plotting the relative number of SSGs, with respect to $N_{\rm stars}$ and $N_{\rm b}$ (second row of Figure~\ref{f:CMC2}), shows no
  significant correlation.

The more interesting result from this comparison is that the number of SSGs produced through both binary channels
increases toward decreasing central density ($\log_{10}(\rho)$), increasing core radii ($r_{\rm c}$) and a decreasing ratio of the 
half-mass to core radii ($r_{\rm h}/r_{\rm c}$).
In other words, these model predicts that diffuse clusters are most efficient at producing SSGs through binary channels.
Furthermore, these trends are far more dramatic for SSGs produced through ongoing mass transfer (``SG MT'').
While our analytic calculations from Section~\ref{s:pchan} only account for disruptions of soft binaries,
the Monte Carlo model predicts that even these hard binaries can be subjected to perturbations, exchanges, etc.\ that can
stop binaries from forming SSGs.  Apparently, the mass transfer channel is particularly vulnerable to these dynamical interruptions \citep[see also][]{leigh:16}.

We also investigate the relation between the number of SSGs and the core collision rate ($\Gamma_{\rm c}$ ;
here we calculate the combined rate for 1+2 and 2+2 encounters, for a binary semi-major axis equal to the 
Roche radius of a 10 Gyr star at the end of the subgiant phase with a 0.45\Msolar\ MS star companion, roughly the expected mean MS mass).
For both the ``SG MT'' and ``SG Mag'' channels, the number of SSGs rises toward modest $\Gamma_{\rm c}$ values ($\sim$0.03 Myr$^{-1}$).
The ``SG MT'' channel then decreases again toward high $\Gamma_{\rm c}$ values, while the ``SG Mag'' channel remains roughly constant.

Generally, as $\Gamma_{\rm c}$ increases, the more frequent dynamical encounters become more efficient at hardening
(i.e., shrinking the semi-major-axis of) hard binaries, in this case to potentially create SSGs through both binary channels.
Additionally, as $\Gamma_{\rm c}$ increases, dynamical exchanges that insert subgiants into sufficiently short-period binaries becomes more likely.
This may account for the increase in the number of SSGs, in both binary channels, up to modest $\Gamma_{\rm c}$ values.

On the other hand, toward higher $\Gamma_{\rm c}$ values, encounters may be energetic and frequent enough to perturb binaries
away from producing SSGs (e.g., through inducing binary coalescence, or otherwise inhibiting mass transfer).
This may, at least partly, explain the decrease in $N_\text{SG MT}$ and the flattening in $N_\text{SG Mag}$ toward
higher $\Gamma_{\rm c}$ values.  Though, we also believe that initial condition choices may contribute to this trend.

Some additional insight into this relation between $\Gamma_{\rm c}$ and the number of SSGs can be found by comparing against the
number of BSS, $N_{\rm BSS}$.
BSS are produced in the Monte Carlo model through both collisions and binary evolution, and here we include both channels in $N_{\rm BSS}$.
For the few models that produce $>$150 BSS 
(beyond the peak in the relation between $N_{\rm BSS}$ and $N_\text{SG MT}$), the mean encounter rate, $\left<\Gamma_{\rm c}\right> \sim 0.19$,
as compared to $\left<\Gamma_{\rm c}\right>\sim0.06$ for models with $<$150 BSS.
At the low $N_\text{BSS}$ and low $\Gamma_{\rm c}$ end, both the SSGs and BSS are produced primarily through binary evolution,
and therefore the number of SSGs increases with increasing number of BSS.
However, the models with high $\Gamma_{\rm c}$ produce BSS primarily through collisions, due to higher encounter rates.
Encounters can also perturb the ``proto - SG MT'' binaries away from producing SSGs through mass transfer, which results in a peaked
distribution of $N_{\rm BSS}$ and $N_\text{SG MT}$.
On the other hand, we see again that the ``SG Mag'' channel is less affected by dynamics, and $N_\text{SG Mag}$ simply continues to increase with $N_\text{BSS}$.

For both the ``SG Mag'' and ``SG MT'' channels, we see the relative number of SSGs with respect to $N_\text{BSS}$ decreases toward larger $N_\text{BSS}$.
Again, the models that produce the most BSS do so primarily through collisions; thus the most interesting portion of this panel is toward the low-$N_\text{BSS}$
end, where the BSS are produced more often through binary evolution (like the SSGs here).  The models predict that for some clusters with low encounter rates,
the number of SSGs may be comparable (to within a factor of a few) to the number of BSS.
  
In summary, the prediction from these Monte Carlo models is that the binary evolution channels dominate the production of SSGs.
Furthermore, the largest number of SSGs produced through the binary evolution channels 
should be found in massive, diffuse clusters with high binary frequencies and modest encounter rates. 
At present, the observed data are too sparse to search for a trend in number of SSGs with encounter rate.  
Nonetheless, this result from the Monte Carlo models aligns with our suggestion in Paper I that dynamical disruptions, perturbations, and 
other alterations to ``proto-SSG'' binaries could 
explain the empirical trend of decreasing specific SSG frequency with increasing cluster mass (Figure~\ref{f:NvM}).  
These dynamical effects inhibit the binary evolution channels, and particularly the ``SG MT'' channel, in clusters
with higher encounter rates (like those in our observed sample of globular clusters).
Clusters with the highest encounter rates may begin to produce SSGs through the collision mechanisms
at a similar, or perhaps higher, rate than the binary mechanisms.

\vspace{2em}
\section{Discussion and Conclusions}
\label{s:discuss}

In Paper I, we identify from the literature a sample of 65 SSG and RS stars in 16 star clusters, including both open and globular clusters, 
and we summarize their empirical demographics within this paper in Section~\ref{s:intro}.
In Paper II, we discuss in detail three potential formation channels for SSGs. The mechanisms within these channels involve isolated subgiant binary evolution,
rapid partial stripping of a subgiant’s envelope (for which we envision two mechanisms, one through common-envelope evolution and another through dynamical encounters),
or reduced luminosity due to magnetic fields that inhibit convection. In addition Paper II briefly considers a formation channel through collisions of two
main-sequence stars during a binary encounter, which we elaborate upon here.

With isolated binaries, SSGs may be produced through ongoing binary mass transfer involving a subgiant star (Section~\ref{s:SGMT}, ``SG MT''), reduced 
convective efficiency on a rapidly rotating magnetically active subgiant, likely in a tidally locked binary (Section~\ref{s:SGMag}, ``SG Mag''), or 
rapid stripping of a subgiant's envelope during a common-envelope phase (Section~\ref{s:SGStrip}, ``SG CE'').  Invoking stellar collisions
(most likely involving at least one binary, \citealt{leigh:12b,leigh:13b}), SSGs can be created through a collision and subsequent merger of two MS stars observed
while contracting back onto the MS (Section~\ref{s:MSColl} ``MS Coll''), or a grazing collision involving a subgiant that rapidly strips much of its envelope 
(Section~\ref{s:SGStrip}, ``SG Coll'').  
The binary evolution channels can happen in isolation, while the collision channels require the dynamical environment of a star cluster.  
Yet all of these channels are catalyzed by binary stars.  

Our analytic Poisson probability calculations (Sections~\ref{s:pchan} and~\ref{s:prob}, which are upper limits)
and our analysis of a large grid of Monte Carlo models (Section~\ref{s:nbody}) suggest that the binary evolution channels are dominant.
In particular, both of these methods predict that we are most likely to observe SSGs that originate from 
magnetically active subgiants with reduced convective efficiency (see Figures~\ref{f:poisson} and~\ref{f:bar}).

This result is based on the SSG formation rates alone, without any constraint on the expected binarity of the product.  
Observationally, we know that the SSGs are primarily in short-period active binaries (Paper I, and see Section~\ref{s:intro} here).
At least two thirds of the SSGs have photometric and/or radial-velocity periods of $\lesssim$15 days, and 
at least three quarters of these variables are confirmed to be radial-velocity binaries.
These short orbital periods are consistent with tidally locked binaries 
\citep[e.g.][]{meibom:05}, as expected for the ``SG Mag'' mechanism.
The SSGs with the shortest-period variability may be in binaries currently (or very recently) undergoing mass transfer.  
Indeed, there are a few W UMa contact binaries amongst the SSGs in our sample (in NGC 188, $\omega$~Centauri and NGC 6397), which support the ``SG MT'' mechanism.
In short, the ``SG MT'' and ``SG Mag'' mechanisms naturally explain the binarity.

Additional empirical evidence supporting SSG formation through isolated binary evolution may be found in the
nearly 10,000 stars in the ``No-Man's-Land'' from Kepler \citep{batalha:13,huber:14}, which may be field SSGs.
These stars are important targets for future observations, and we will investigate them in more detail within a future paper.  

Conversely, producing SSGs through collisions may only be relevant in very dense star clusters.
Furthermore, encounters that lead to the ``MS Coll'' mechanism generally produce collision products in wider binaries (or without companions),
sometimes with periods that are orders of magnitude larger than observed for the SSGs \citep{leigh:11b,geller:13}.
When also considering the low Poisson probabilities calculated here for the ``MS Coll'' channel,
and the even lower number predicted by the Monte Carlo models (see Figure~\ref{f:poisson} and Section~\ref{s:nbody})
we conclude that, in most clusters, observing a SSG from the ``MS Coll'' channel is unlikely, especially for SSGs found in a short-period binary.
The few globular clusters studied in Paper I with very high encounter rates may be the best places to find SSGs produced through this mechanism (see Section~\ref{s:pclu} 
and Table~\ref{probtab}).

Observing a SSG resulting from the rapid loss of a subgiant's envelope (``SG Strip''), through either mechanism explored here, is also relatively unlikely, 
given our Poisson probability calculations and our analysis of the Monte Carlo models.
The expected binarity of the product for ``SG Strip'' is less clear than for the other mechanisms.
It may be possible that a grazing encounter that strips a subgiant's envelope can leave a bound companion in a short-period binary (akin to a tidal capture binary), 
but further study is required to confirm if this is indeed possible.  
Likewise stripping in common-envelope evolution is highly uncertain, and it is unclear what the binarity of the product would be. 

Other efficient mechanisms may also exist that we have not identified, which could explain why our $\Psi(n_\text{SSG})$ Poisson probabilities do not reach unity for 
some clusters (and particularly the open clusters) in Table~\ref{probtab}, where $n_\text{SSG}$ SSGs are in fact observed.  
For instance, there may be other ``SG Strip'' mechanisms that we have not investigated.
Perhaps SSGs can be created if stable mass transfer is interrupted dynamically, as discussed in \citet{leigh:16}.
In addition, very close companions to neutron stars can be evaporated, as in the well-known ``black widow'' pulsars \citep[e.g.][]{fruchter:90}.  
Perhaps companions in the early stages of being evaporated would appear as SSGs, as may be the case for SSG U12 in NGC 6397 \citep{damico:01,ferraro:03}.

Massive and diffuse globular clusters may be the most promising targets for future observations aimed at identifying additional SSGs.
The Monte Carlo globular cluster models (Section~\ref{s:nbody}) predict that such clusters should have the largest frequency of SSGs 
created through the binary evolution channels.  
The Monte Carlo models also predict that the binary evolution channels may be inhibited for the densest clusters with high encounter rates, which is
consistent with the current observations (Figure~\ref{f:NvM}; though note that the observations are incomplete, see Paper I).
It is clear that in some clusters multiple mechanisms likely operate simultaneously to produce SSGs (e.g., see Table~\ref{probtab}).

Many of these observed and predicted trends in number of SSGs are also seen for BSS.  For instance, the frequency of BSS in globular clusters
is observed to be anticorrelated with the absolute luminosity (mass) of the cluster \citep{piotto:04, leigh:07}, but correlated with the
binary fraction \citep{sollima:08, milone:12}. These observations point to binaries as a critical ingredient for BSS formation in
globular clusters \citep{knigge:09}.
The correlations seen in globular cluster observations have been interpreted theoretically to indicate that binary evolution is an important, and sometimes dominant,
BSS production mechanism \citep{leigh:11b}, though binary-mediated collisions may also be important at high densities \citep{sills:13,chatterjee:13a}.
The reduced survival of binaries (i.e., BSS and SSG progenitors) in high density (and high velocity dispersion)
environments likely also contributes to these observed correlations \citep{davies:04, sollima:08b}, as does the preferential retention of binary stars, compared
to the less massive single stars, in clusters that experience significant mass loss (as may be the case for the lower-mass clusters in our observed SSG sample).
Binaries are also critical for BSS (and likely also SSG) formation in open clusters \citep{mathieu:09} and the field \citep{carney:05}.
The discussion from this body of literature may help to explain the observed decreasing trend in specific frequency of
SSGs with increasing cluster mass, shown in Figure~\ref{f:NvM}.

Though we focus on the SSGs throughout the majority of the paper, the RS stars (i.e, stars that occupy the lighter gray regions in 
Figure~\ref{f:CMDreg}) have very similar empirical characteristics (Paper I).
As shown in Figure~\ref{f:CMDreg}, RS and SSG stars may be produced through the same mechanisms, and in some cases one can be the evolutionary precursor to the other.
Furthermore, at least two of these mechanisms that form SSGs, mass transfer and MS -- MS collisions, are also invoked to explain the origins of BSS 
and yellow stragglers/giants \citep{mccrea:64,mathieu:86b,leonard:89,chen:08,leigh:11b,chatterjee:13a,sills:13,gosnell:15,leiner:16}.
Some fraction of these stars may have been born through the same (or similar) formation channels, and perhaps in some cases these stars may represent 
different stages along the same evolutionary sequence.
Comparing the frequencies and binary characteristics of these stellar populations across multiple star clusters could reveal important insights into 
their formation mechanism(s) and provide important guidance for detailed evolutionary models of binary mass transfer and the products of stellar collisions.

\acknowledgments
A.M.G.\ acknowledges support from NASA through HST grant AR-13910 and a National Science Foundation Astronomy and Astrophysics Postdoctoral Fellowship 
Award No.\ AST-1302765.
S.C.\ acknowledges support from NASA through HST grant HST-AR-12829.004-A.
Support for Programs AR-13910 and HST-AR-12829.004-A were provided by NASA through a grant from the Space Telescope Science Institute,
which is operated by the Association of Universities for Research in Astronomy, Incorporated, under NASA contract NAS5-26555.
This research was supported in part through the computational resources and staff contributions provided for the Quest high performance 
computing facility at Northwestern University which is jointly supported by the Office of the Provost, the Office for Research, and 
Northwestern University Information Technology.

\bibliographystyle{aasjournals}
\bibliography{SSG_frequencies}

\clearpage
\LongTables
\begin{landscape}

\setlength{\tabcolsep}{4pt}
\begin{deluxetable}{lcccccccccccccccc}
\tabletypesize{\tiny}
\tablecaption{Sub-subgiant Formation Probabilities \label{probtab}}
\tablehead{\colhead{Cluster} & \colhead{age} & \colhead{[Fe/H]} & \colhead{$M_{\rm cl}$} & \colhead{$f_{\rm b}$} & \colhead{$\sigma_0$} & \colhead{$\log(\rho_0)$} & \colhead{$r_{\rm c}$} & \colhead{$r_{\rm hm}$} & \colhead{$P_{\rm circ}$} & \colhead{$n_{\rm SSG}$} & \colhead{$\Psi_{\rm SG\ MT}$} & \colhead{$\Psi_{\rm MS\ Coll}$} & \colhead{$\Psi_{\rm SG\ Coll}$} & \colhead{$\Psi_{\rm SG\ CE}$} & \colhead{$\Psi_{\rm SG\ Mag}$} & \colhead{$\Psi(n_{\rm SSG})$}  \\ 
\colhead{} & \colhead{[Gyr]} & \colhead{} & \colhead{[M$_\odot$]} & \colhead{} & \colhead{[km s$^{-1}$]} & \colhead{[M$_\odot$/pc$^3$]} & \colhead{[pc]} & \colhead{[pc]} & \colhead{[day]} & \colhead{} & \colhead{} & \colhead{} & \colhead{} & \colhead{} & \colhead{} & \colhead{} }
\tablewidth{0pt}
\startdata
&&&&&&&&&&&&&&\\
\multicolumn{15}{l}{\textbf{Open Clusters}} \\
&&&&&&&&&&&&&&\\
NGC 188    &               6.2   &             0.0   &              1500$\pm$400   &    0.5$\pm$0.05   &   0.41$\pm$0.04   &           \nodata   &               2.1   &               4.0   &    14.5$\pm$1.8   &  3   &                                            $0.042$   &                                            $0.003$   &                                             $0.02$   &                                             $0.02$   &                                              $0.5$   &                                             $0.04$ \\
NGC 2158   &                 2   &            -0.6   &                     15000   &         \nodata   &         \nodata   &           \nodata   &              3.23   &           \nodata   &         \nodata   &  1   &                                             $0.05$   &                                            $0.006$   &                                            $0.001$   &                                           $0.026$   &                                             $0.27$   &                                             $0.33$ \\
NGC 2682   &                 4   &             0.0   &              2100$\pm$600   &   0.57$\pm$0.04   &   0.59$\pm$0.07   &           \nodata   &                 1   &           \nodata   &    12.1$\pm$1.3   &  2   &                                             $0.05$   &                                             $0.05$   &                                              $0.1$   &                                              $0.0$   &                                             $0.47$   &                                              $0.2$ \\
NGC 6791   &                 8   &             0.4   &             4600$\pm$1500   &         \nodata   &    0.62$\pm$0.1   &           \nodata   &               3.4   &           \nodata   &         \nodata   &  5   &                                            $0.086$   &                                            $0.004$   &                                             $0.03$   &                                             $0.04$   &                                             $0.82$   &                                             $0.04$ \\
NGC 6819   &               2.4   &             0.0   &                      2600   &    0.4$\pm$0.02   &         \nodata   &           \nodata   &              1.75   &           \nodata   &     6.2$\pm$1.1   &  1   &                                            $0.016$   &                                            $0.005$   &                                            $0.001$   &                                            $0.008$   &                                             $0.12$   &                                             $0.15$ \\
NGC 7142   &               3.6   &             0.1   &                       500   &         \nodata   &         \nodata   &           \nodata   &               3.1   &           \nodata   &         \nodata   &  0   &                                            $0.009$   &                                          $0.00013$   &                                           $0.0003$   &                                           $0.0045$   &                                              $0.1$   &                                            \nodata \\
&&&&&&&&&&&&&&\\
\multicolumn{15}{l}{\textbf{Globular Clusters}} \\
&&&&&&&&&&&&&&\\
NGC 104    &            13.1   &           -0.72   &      1.0$\times 10^{6}$   &   0.02$\pm$0.01   &      11$\pm$0.3   &             5.18   &             0.47   &             4.15   &         \nodata   &  8   &                                             $0.71$   &                                                $1$   &                                                $1$   &                                            $0.46$   &                                                $1$   &                                                $1$ \\
NGC 5139   &            11.5   &           -1.53   &      2.2$\times 10^{6}$   &         \nodata   &    16.8$\pm$0.3   &             3.45   &             3.59   &             7.56   &         \nodata   & 15   &                                            $0.99$   &                                              $0.4$   &                                              $0.4$   &                                              $0.9$   &                                                $1$   &                                                $1$ \\
NGC 6121   &            12.5   &           -1.16   &      1.3$\times 10^{5}$   &    0.1$\pm$0.01   &       4$\pm$0.2   &             3.94   &             0.74   &             2.77   &         \nodata   &  2   &                                              $0.4$   &                                             $0.73$   &                                             $0.85$   &                                             $0.25$   &                                                $1$   &                                                $1$ \\
NGC 6218   &            12.7   &           -1.37   &      1.4$\times 10^{5}$   &   0.06$\pm$0.01   &     4.5$\pm$0.4   &             3.53   &             1.10   &             2.47   &         \nodata   &  1   &                                             $0.38$   &                                              $0.3$   &                                              $0.4$   &                                              $0.2$   &                                                $1$   &                                                $1$ \\
NGC 6366   &            13.3   &           -0.59   &      4.8$\times 10^{5}$   &   0.11$\pm$0.03   &     1.3$\pm$0.5   &             2.70   &             2.21   &             2.98   &         \nodata   &  1   &                                              $0.6$   &                                              $0.2$   &                                              $0.4$   &                                              $0.4$   &                                                $1$   &                                                $1$ \\
NGC 6397   &            12.7   &           -2.02   &      7.7$\times 10^{4}$   &   0.02$\pm$0.01   &     4.5$\pm$0.2   &             6.06   &             0.03   &             1.94   &         \nodata   &  3   &                                             $0.07$   &                                              $0.5$   &                                              $0.3$   &                                              $0.0$   &                                              $0.5$   &                                              $0.3$ \\
NGC 6652   &            12.9   &           -0.81   &      7.9$\times 10^{4}$   &    0.1$\pm$0.01   &         \nodata   &             4.78   &             0.29   &             1.40   &         \nodata   &  0   &                                             $0.89$   &                                             $0.89$   &                                                $1$   &                                          $0.68$   &                                                $1$   &                                            \nodata \\
NGC 6752   &            11.8   &           -1.54   &      2.1$\times 10^{5}$   &   0.01$\pm$0.01   &     4.9$\pm$0.4   &             5.34   &             0.20   &             2.22   &         \nodata   &  0   &                                              $0.1$   &                                             $0.97$   &                                             $0.97$   &                                             $0.05$   &                                              $0.6$   &                                            \nodata \\
NGC 6809   &            12.3   &           -1.94   &      1.8$\times 10^{5}$   &         \nodata   &       4$\pm$0.3   &             2.52   &             2.83   &             4.45   &         \nodata   &  2   &                                             $0.43$   &                                             $0.07$   &                                             $0.04$   &                                             $0.25$   &                                             $0.99$   &                                            $0.98$ \\
NGC 6838   &            12.0   &           -0.78   &      3.0$\times 10^{4}$   &   0.22$\pm$0.02   &     2.3$\pm$0.2   &             3.13   &             0.73   &             1.94   &         \nodata   &  2   &                                             $0.33$   &                                             $0.13$   &                                              $0.4$   &                                              $0.2$   &                                                $1$   &                                                $1$ \\
\enddata
\tablenotetext{}{Note: References for the values in this table, other than the probabilities, are as follows. For the \textbf{open clusters}: 
\textit{NGC 188:} We take the age from \citet{meibom:09} and the adpoted [Fe/H] from \citet{sarajedini:99}, $M_{\rm cl}$, $r_{\rm c}$, $r_{\rm hm}$ from \citet{chumak:10}, $f_{\rm b}$ from \citet{geller:13}, $\sigma_0$ from \citet{geller:08} and $P_{\rm circ}$ from \citet{meibom:05}.
\textit{NGC 2158:} We take the age,  [Fe/H], $M_{\rm cl}$ from \citet{carraro:02}, and $r_{\rm c}$ from \citet{kharchenko:13}.
\textit{NGC 2682:} We take the age,  [Fe/H], $M_{\rm cl}$, $f_{\rm b}$, $\sigma_0$, $r_{\rm c}$ from \citet[][and references therein]{geller:15}, and  $P_{\rm circ}$ from \citet{meibom:05}.
\textit{NGC 6791:} We take the age,  [Fe/H] from \citet{carney:05}, $M_{\rm cl}$, $\sigma_0$ from \citet{tofflemire:14}, and $r_{\rm c}$ from \citet{platais:11}.
\textit{NGC 6819:} We adopt the age,  [Fe/H] from \citet[][and references therein]{hole:09}, take  $M_{\rm cl}$, $r_{\rm c}$ from \citet{kalirai:01}, and $f_{\rm b}$ (scaled here to full period distribution using method from \citealt{geller:15}), $P_{\rm circ}$ from \citet{milliman:14}.
\textit{NGC 7142:} We take the age,  [Fe/H] from \citet[][and references therein]{sandquist:13}, estimate $M_{\rm cl}$ from \citet{straizys:14}, and take $r_{\rm c}$ from \citet{kharchenko:13}.
For the \textbf{globular clusters} we take the age from \citet[][using the ``G00$_{\rm CG}$'' values and normalized using the age of 47 Tuc from \citealt{thompson:10}]{marin:09}, [Fe/H], $\sigma_0$ (where available), $M_{\rm cl}$ and $\log{\rho_0}$ (both calculated assuming a mass-to-light ratio of 2), $r_{\rm c}$, $r_{\rm hm}$ from \citet{harris:96,harris:10}, and $f_b$ (where available) from \citet{milone:12}. For \textit{NGC 6366} we calculate the mass from $\sigma_0$ assuming a Plummer model.  Finally for \textit{NGC 6838} we take the age from \citet{dicecco:15}.
}

\end{deluxetable}

\setlength{\tabcolsep}{6pt}

\clearpage
\end{landscape}

\end{document}